%% file: main_text.tex
\documentclass[a4paper]{ar-1col}

\usepackage{url}
\usepackage{hyperref}
\hypersetup{colorlinks,linkcolor=red,citecolor=blue,urlcolor=blue}

\usepackage[svgnames]{xcolor}
\definecolor{DarkOrange}{rgb}{1, 0.35, 0}

\newcommand{\teff}{$T_{\mathrm{eff}}$}
\newcommand{\logg}{$\log g$}
\newcommand{\feh}{[Fe/H]}
\newcommand{\afe}{[$\alpha$/Fe]}
\newcommand{\xfe}{[X/Fe]}
\newcommand{\mh}{[M/H]}
\newcommand{\vmic}{$v_{\mathrm{mic}}$}
\newcommand{\vmac}{$v_{\mathrm{mac}}$}

\newcommand{\gbs}{{\it Gaia} benchmark star}
\input{species}

\usepackage{amssymb}
\usepackage[normalem]{ulem}

\usepackage{natbib}
\bibpunct{(}{)}{;}{a}{}{,}
\input{journals}
\jname{Xxxx. Xxx. Xxx. Xxx.}
\jvol{AA}
\jyear{YYYY}
\doi{10.1146/((please add article doi))}

\begin{document}

\markboth{Jofr\'e et al.}{Industrial stellar abundances}

\title{Accuracy and precision of industrial stellar abundances}


\author{Paula Jofr\'e,$^1$ Ulrike Heiter,$^2$ and Caroline Soubiran$^3$
\affil{$^1$N\'ucleo de Astronom\'ia, Facultad de Ingenier\'ia y Ciencias, Universidad Diego Portales, Av. Ej\'ercito 441, Santiago, Chile; email: paula.jofre@mail.udp.cl}
\affil{$^2$Observational Astrophysics, Department of Physics and Astronomy, Uppsala University, Box 516, 75120 Uppsala, Sweden}
\affil{$^3$Laboratoire d'Astrophysique de Bordeaux, Univ. Bordeaux, CNRS, B18N, all\'ee Geoffroy Saint-Hilaire, F-33615, Pessac, France}}

\begin{abstract}
There has been an incredibly large investment in obtaining high-resolution stellar spectra for determining chemical abundances of stars. This information is crucial to answer fundamental questions in Astronomy by constraining the formation and evolution scenarios of the Milky Way as well as the stars and planets residing in it.

We have just entered a new era, in which chemical abundances of FGK-type stars are being produced at industrial scales, where the observations, reduction, and analysis of the data are automatically performed by machines. Here we review the latest human efforts to assess the accuracy and precision of such industrial abundances by providing insights in the steps and uncertainties associated with the process of determining stellar abundances.

To do so, we  highlight key issues in the process of spectral analysis for abundance determination, with special effort in disentangling sources of uncertainties. We also provide a description of current and forthcoming spectroscopic surveys, focusing on their reported abundances and uncertainties. This allows us to identify which elements and spectral lines are best and why. Finally, we make a brief selection of main scientific questions the community is aiming to answer with abundances. 

\end{abstract}

\maketitle

\tableofcontents

\section{INTRODUCTION}

The elemental abundances of FGK-type stars provide key pieces of information for characterising the stellar populations of our Galaxy. Different stellar populations have different chemical patterns, and the foundation for explaining these differences is well-established: chemical elements are created in a variety of nucleosynthesis channels inside stars, and are distributed into the Galaxy either through supernovae or stellar winds. New stars are born from this enriched material, creating new elements which are then sent back to the interstellar medium. This cycle has been repeating ever since the formation of the first stars until today.

The outcome of Galactic chemical evolution is more complex than what is implied by the simple description above, considering the variety of stellar masses and therefore lifetimes, and the diversity of physical processes taking place inside stars. 
Therefore, accurate and precise abundances of large samples of stars are required to constrain chemical evolution models. The productions of elements (yields) are different for stars at different masses and metallicities; the amount of the enriched material recycled into new stars depends on the total mass of the Galaxy because it must be able to keep the gas bound to form new stars. Since the mases and the sizes of galaxies change with time, so does the star formation rate and the subsequent chemical enrichment. Finally, we know that galaxies experience inflow and outflow of material due to, e.g., accretion of other galaxies, which have  different chemical enrichment histories and stellar populations with other chemical patterns \citep[see e.g][for a description of the ingredients in chemical evolution models]{2006ApJ...653.1145K}. FGK-type stars live long enough and have shallow convective zones, so that the information on the chemical make-up of the gas from which they formed is retained in their spectra. Hence, their abundances are the best fossil records we can use to constrain the cosmic matter cycle. However, these fossils move about in the Galaxy. With the help of a dynamical model and the ages of the stars it might be possible to find their original site of formation \citep{2002ARA&A..40..487F}. Then the fossils of a stellar population might be found, and the ingredients of its chemical evolution constrained \citep[and references therein]{2013NewAR..57...80F}.

First works putting these pieces together were limited by the lack of good measurements of distances which did not allow probing the distribution of chemical elements in the Galaxy. Even if stellar abundances were believed to be of reasonable accuracy, it was not possible to constrain a chemodynamical model with the scarcity of data on distances, kinematics and ages. However, these very struggling scientists provided the motivation for the projects that are responsible for the wealth of stellar data we have today, starting from the revolutionary Gaia mission \citep{2018A&A...616A...1G}, followed by the large spectroscopic and asteroseismic surveys. This industrial revolution in Galactic astronomy is only beginning, as more data releases of Gaia are approaching, and more spectroscopic and seismic surveys are planned.

Newer generations of scientists have the opportunity to work with these ready-to-use data products. 
Now that the major challenge of good measurements of distances is largely solved thanks to Gaia, do we believe that stellar abundances are of sufficient accuracy? High resolution multi-object spectrographs are restricted to point towards the sky each from a different spot on the surface of the Earth, with different instruments. It is natural that the data products from different surveys will differ, but how is that limiting our capacity to unravel the structure and formation of our Galaxy?
This review intends to answer some of these questions, starting with an overview of the major steps involved in the derivation of stellar abundances in Sect.~\ref{sect:methods}, followed by Sect.~\ref{sect:errors} suggesting standard ways to quantify the uncertainties in the results. In Sect.~\ref{sect:elements}, we summarise the large datasets with abundances available today, discussing why we know more about some elements than others. We continue with a review on the progress the field has experienced thanks to stellar abundances in Sect.~\ref{sect:science}, and finish with a discussion answering these questions and some thoughts on the future in Sect.~\ref{sect:discussion}.

\section{FROM SPECTRA TO ABUNDANCES: STEPS AND ISSUES}
\label{sect:methods}
\input{sect_methods}

\section{ASSESSING THE ABUNDANCE ERROR BUDGET}
\label{sect:errors}


\input{uncertainties}

\section{THE PERIODIC TABLE AS SEEN FROM SPECTRAL ANALYSES}
\label{sect:elements}

We start with an overview of relatively large ($\sim$1000 stars) catalogues of stellar abundances, followed by how they have served to build the industrial products from spectroscopic surveys available today and in the future. That information will then help us to understand why certain elements are more popular than others, as well as to discuss how different surveys compare for common stars. 

\input{surveys}
\label{sect:survey_description}

\subsection{Discussion on individual elements}
\label{sect:common_elements}

\input{sect_common_elements}


\section{RECONSTRUCTING THE HISTORY OF THE GALAXY WITH ABUNDANCES}

\label{sect:science}

\input{sect_stellarPops}

\section{DISCUSSION AND CONCLUDING REMARKS}
\label{sect:discussion}

Astronomy, being one of the oldest sciences of mankind, has been traditionally hampered by the lack of good data. Astronomers have thus developed the habit to blame the small amount of data available for the unanswered questions. They have been dreaming of having millions of stars with accurate data, and that the wealth of data will help us to progress in understanding how our own Galaxy is shaped. 

This dream is becoming true! Today, Galactic astronomy benefits from exquisite data: billions of accurate parallaxes from Gaia and thousands of high-resolution spectra and asteroseismic data are enabling us to dig into the physics of stars. We are experiencing a unique opportunity to re-formulate our understanding about stellar properties because data quantity and quality are not a problem anymore. The variety of data available naturally has led to a variety of analysis methods, which, at this rapidly growing data rate, have the dangerous potential to diverge significantly and so their results. Fortunately, we are learning that the best way to maximise the accuracy and precision of our data products is to work together as a community, where surveys need to be compared and complemented.  RAVE has provided a crucial playground for learning what we will obtain from the forthcoming millions of similar spectra from Gaia-RVS; the Gaia-ESO survey has been revolutionary in making us become aware of the impact of different methodologies on results; APOGEE has shown us the power of moving outside our wavelength range of comfort (from the optical to the infrared); GALAH is going one step ahead in pushing for better modelling and propagating the improved parameters with data-driven approaches to an entire dataset; and SDSS and LAMOST are demonstrating that we are able to trace the chemistry of huge volumes of the Galaxy even with low resolution. 

We still do not have a best method to determine accurate and precise stellar parameters and abundances for all stars in the Galaxy. But inter-comparisons between surveys are starting to become mandatory thanks to new efforts to observe common targets between surveys. We are getting closer to defining standard procedures to compare and connect results, being more aware of uncertainties and having well-defined strategies to improve them. We are also learning to appreciate that we need both ``high'' and ``low'' quality data. High-quality data (high S/N and resolution) are crucial to improve the theory of line formation, as well as to identify and study the chemical signatures which are needed to maximise the size of the chemical-space in the Galaxy. Seemingly low-quality data (low S/N and resolution) still provide the only realistic way to travel across the Galaxy and probe its outskirts. It is clear that the information we can obtain from the latter kind of data fully depends on what we can obtain from the former. Thus, a concerted effort is key for taking full advantage of the treasure which the previous generation of astronomers has given us thanks to their habit of wanting more data. 

Working together means also to make a serious effort of standardising our data products, not only in terms of meaningful physical properties (e.g.,  \feh, \afe, etc.), but also in terms of their format. Catalogues, in particular those created by independent groups intended to be used for reference, must be published in the Virtual Observatory (via CDS). This one extra step in the publishing procedure can lead to productive synergies in the community. Reference stars are the basis for Galactic studies and must be accessible for the entire community, if we want pipeline products to truly converge in accuracy. Likewise, concluding on the best suitable pipeline to put survey abundances onto one common scale is only possible if the spectra are public for everyone, providing the opportunity to freely test their tools and reproduce results. We should learn from the examples of SDSS, APOGEE, LAMOST, and Gaia. 

This review has attempted to link details of the art of determining abundances of a single line from a single star with the art of propagating this information to millions of stars, and how each step and star is crucial for this chemical ladder to work. In Sect.~\ref{sect:methods}, we described the several steps involved in abundance determination, and in Sect.~\ref{sect:errors}, we discussed different tests that help us to quantify the uncertainties in abundances. In Sect.~\ref{sect:elements}, we described how catalogues and surveys apply these steps to provide abundances at industrial scales, discussing which elements are common and which are not. Finally, in Sect.~\ref{sect:science}, we discussed a selection of science applications where these abundances are being used. 

The Milky Way, our home Galaxy, harbours stars of a great variety, each of them containing unique information about their present and past environment. The evolution of this environment encodes the laws of physics in an elegant way that we are yet to fully decipher, with the fossil stars there to help us out. Starting from the Sun we need to find the best way to connect to other reference stars, ensuring that accuracy and precision are maintained. From these reference stars we can then connect to all other stars of the Milky Way, taking care that we are not overlooking even one single star that might contain the key missing information about the assembly history of our home Galaxy.

\section*{ACKNOWLEDGMENTS}

 We apologise in advance for all the investigations which could not be properly cited here due to space limitations. We thank our collaborators K. Hawkins, P. Das and M. Tucci Maia for various discussions,  S. Buder for clarifying details about the GALAH pipeline and C. Worley, R. Smiljanic and G. Gilmore for clarifying details of the GES pipeline. We further thank P. E. Nissen and B. Gustafsson for fruitful discussions, and for sharing unpublished material which was needed for writing complementary reviews on stellar abundances and T. Beers for his careful revision. P.J. and U.H. thank C. Hidalgo for hosting us at the MIT Media Lab at a critical time of writing this review. Finally, P.J. warmly thanks T. M\"adler for unlimited  support in all matters. 

P.J. acknowledges financial support of FONDECYT Iniciaci\'on 11170174. U.H. acknowledges support from the Swedish National Space Agency (SNSA/Rymdstyrelsen). Posted with permission from the Annual Review of Astronomy \& Astrophysics, Volume 57 \copyright\ by Annual Reviews, \url{http://www.annualreviews.org}.

\input{supplementary_material.tex}

\bibliographystyle{ar-style2}
\bibliography{calibrating_abundances}

\end{document}

%% file: species.tex
\newcommand{\hi}{H\,\textsc{i}}

\newcommand{\ci}{C\,\textsc{i}}
\newcommand{\oi}{O\,\textsc{i}}
\newcommand{\nai}{Na\,\textsc{i}}
\newcommand{\mgi}{Mg\,\textsc{i}}

\newcommand{\ali}{Al\,\textsc{i}}
\newcommand{\sii}{Si\,\textsc{i}}

\newcommand{\cai}{Ca\,\textsc{i}}
\newcommand{\caii}{Ca\,\textsc{ii}}

\newcommand{\scii}{Sc\,\textsc{ii}}
\newcommand{\tii}{Ti\,\textsc{i}}

\newcommand{\vi}{V\,\textsc{i}}

\newcommand{\cri}{Cr\,\textsc{i}}
\newcommand{\crii}{Cr\,\textsc{ii}}

\newcommand{\fei}{Fe\,\textsc{i}}
\newcommand{\feii}{Fe\,\textsc{ii}}

\newcommand{\nii}{Ni\,\textsc{i}}

\newcommand{\gei}{Ge\,\textsc{i}}

\newcommand{\rbi}{Rb\,\textsc{i}}

\newcommand{\zri}{Zr\,\textsc{i}}

\newcommand{\nbi}{Nb\,\textsc{i}}
\newcommand{\moi}{Mo\,\textsc{i}}

\newcommand{\baii}{Ba\,\textsc{ii}}

\newcommand{\ceii}{Ce\,\textsc{ii}}

\newcommand{\ndii}{Nd\,\textsc{ii}}

\newcommand{\ybii}{Yb\,\textsc{ii}}

%% file: sect_methods.tex
\begin{figure}
\hspace{-2cm}
\includegraphics[scale=0.45]{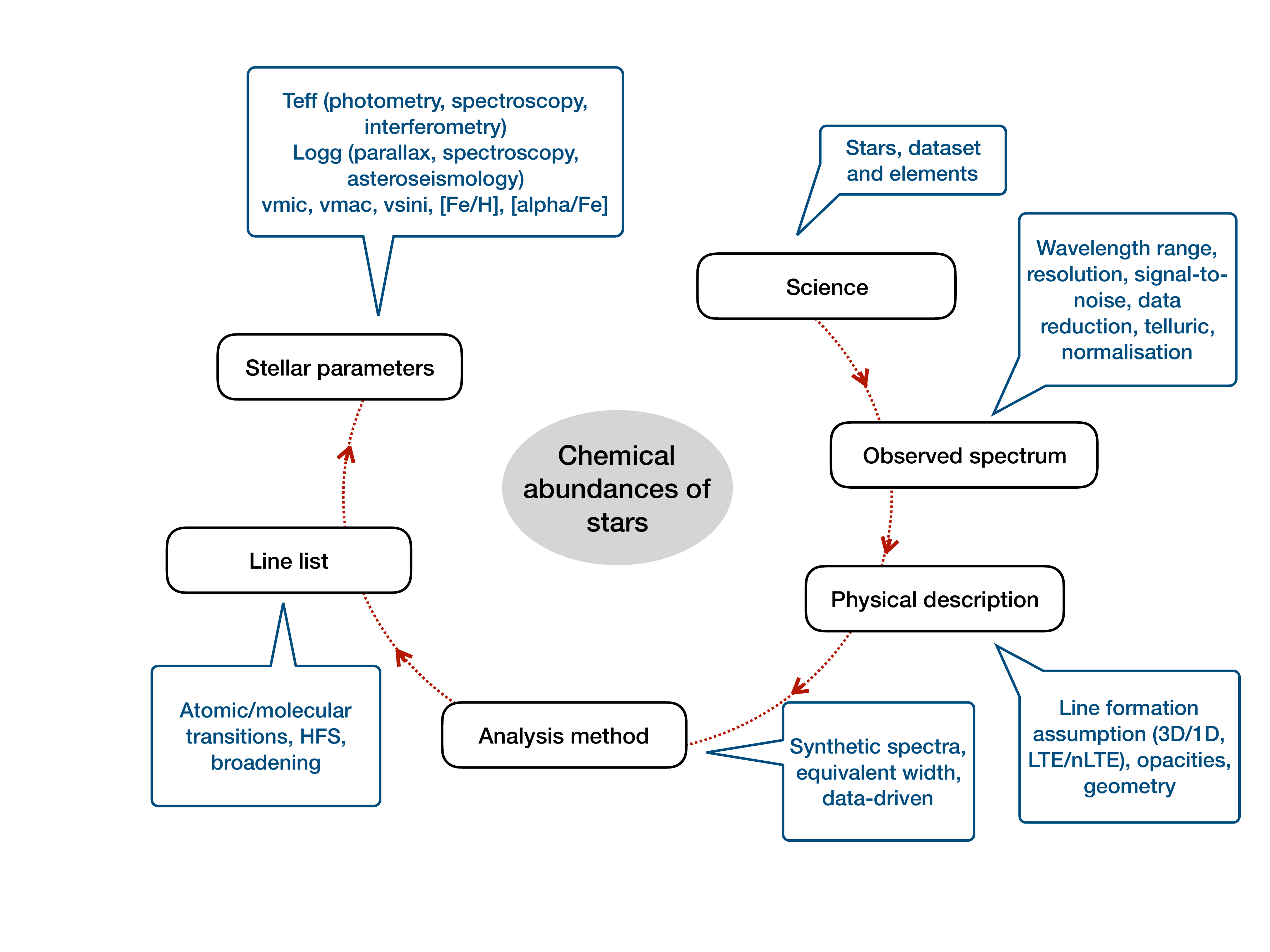}
\caption{Illustration of the steps needed for obtaining abundances of chemical elements in stars. Each of these steps implies uncertainties in the derived abundances which might affect different science cases to different degrees.}
\label{2.method_scheme}
\end{figure}

We complement the brief but comprehensive review of \cite{2016LRSP...13....1A} 
 by illuminating the main steps in the process of deriving stellar abundances  which are illustrated in Figure~\ref{2.method_scheme}, while publicly available tools and material are listed  in   Table~\ref{tab:tools} .

\input{table_tools}

\subsection{Science: selection of stellar sample and chemical elements}
The scientific question will determine the type of stars to study and will dictate the properties of the spectra, in particular their wavelength coverage. 
While in some cases a carefully selected sample of few stars is sufficient to produce revolutionary scientific results \citep[e.g.,][]{1998A&A...338..161F, 2009ApJ...704L..66M, 2010A&A...511L..10N}, 
the increase in computing power and efficiency of data storage has been driving the field to evolve towards a more industrial scale. This is especially the case for studies of the Milky Way structure and evolution, which is the objective of several ongoing large-scale spectroscopic surveys. \cite{2016MNRAS.461.2174R} 
present an interesting discussion of how to define a spectral dataset that will meet these science goals.

\subsection{Observed Spectra}


\paragraph*{Resolution, signal-to-noise, and time dependencies:}

\begin{figure}
\hspace{-2cm}
\includegraphics[scale=0.27,trim=200 0 175 0,clip]{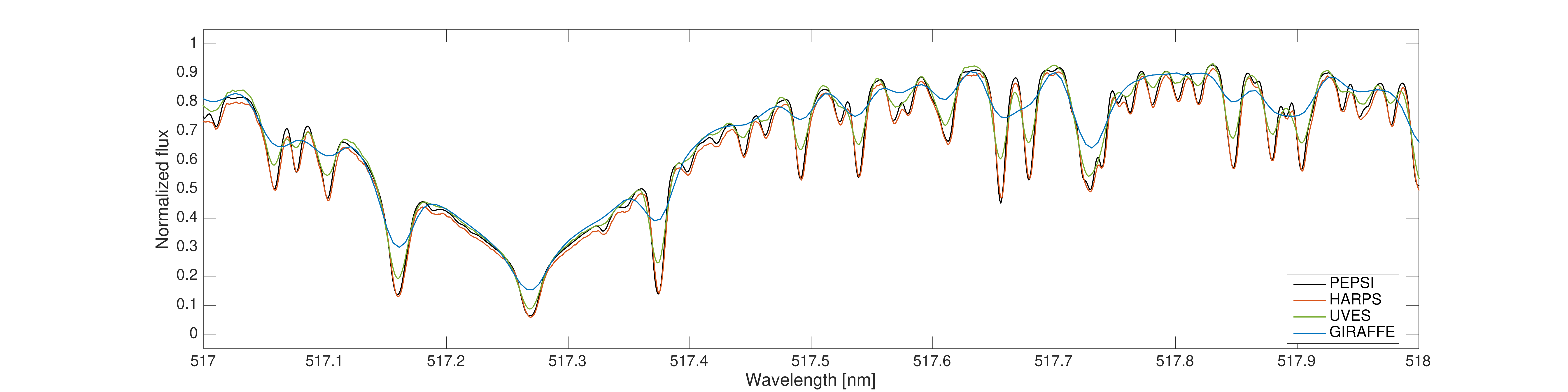}
\caption{Spectra for the \gbs\ $\epsilon$~Eri near one of the \mgi~b lines, obtained with the PEPSI, HARPS, UVES, and GIRAFFE spectrographs.  See text for details}
\label{fig:resolution}
\end{figure}

Spectral lines will be resolved if the instrumental broadening is less than the broadening mechanisms in the stellar atmosphere caused primarily by Doppler broadening due to temperature, turbulence, and rotation \citep[see][Sect.~2]{nissen-gustafsson}.  If many stars need to be analysed, the instrumental resolution does not need to be much higher than that corresponding to the intrinsic stellar one for the purpose of determining abundances as this saves significant observing time. However, a higher resolution will allow one to investigate effects on line profiles such as star spots, asymmetries due to convection, variations due to non-radial pulsations, or blends. Figure~\ref{fig:resolution} compares spectra with different resolutions for  $\epsilon$~Eri.   
The resolving power, S/N, and reference for each spectrum are, respectively,
$\sim$25\,900/330/Gaia-ESO DR3\footnote{\url{http://archive.eso.org/wdb/wdb/adp/phase3_main/form?collection_name=GAIAESO&release_name=DR3}} for GIRAFFE,
$\sim$47\,000/173/Gaia-ESO DR3 for UVES,
$\sim$115\,000/474/\citet{GBS2} 
for HARPS, and
$\sim$200\,000/1350/\citet{2018A&A...612A..45S} 
for PEPSI.

Figure~\ref{fig:resolution} shows how the number of spectral features grows with increasing spectral resolution.  For example, at around 517.9~nm a broad feature is visible in the GIRAFFE spectrum, which is resolved into two components in the UVES spectrum, while the HARPS and PEPSI spectra show a blend of at least four lines. These are identified as being due to \tii, \fei, \vi, \nii, and several MgH lines by comparison with a synthetic spectrum.

A S/N $>$200 might be desirable to determine abundances with high confidence, but in general a S/N of 60--100 seems to be good enough for pipelines to derive accurate abundances of most elements for FGK-type Population I and II stars. Below S/N$\sim$40 abundances become more uncertain, and below 20 they are usually considered to be unreliable \citep[see, e.g.,][]{2014A&A...561A..93H,2014A&A...570A.122S}. 
The latter may be the case for a considerable fraction of current spectroscopic surveys (see Sect.~\ref{sect:surveys}).  Spectra of very low resolution or S/N rarely are able to provide information for a large variety of abundances, although new techniques implementing machine-learning approaches seem promising \citep{2018PASA...35....3N, 2018arXiv180804428L}. 
They still rely on a training set for which abundances are known from more classical methods calibrated with high-resolution and high S/N spectra. It is important to keep in mind that S/N depends on wavelength, where the blue part of the spectrum often has much lower S/N than the red part.

\paragraph*{Data reduction issues:}

Modern high-resolution spectra are commonly taken with cross-dispersed echelle spectrographs, which provide efficient access to an extended wavelength coverage. However, the reduction and extraction of the science spectrum from such images is challenging. %
A seemingly obvious requirement for science-ready spectra is a proper wavelength calibration, but recurring cases of failure or inaccuracies in wavelength calibration have been reported, which may affect abundance analyses \cite[e.g.,][their Sect.~5.2]{2016ApJS..226....4H}, 
The central wavelengths of spectral features need to be accurate enough to achieve a match with the line list. Automatic procedures need to be able to deal with an imperfect wavelength solution by applying a wavelength dependent radial-velocity correction or a sufficiently wide search window for identifying lines.

Merging the orders is another major challenge in extracting echelle spectra. 
\cite{2002A&A...385.1095P}  
provide a detailed description of this process. Orders are curved, thus deviating from the straight lines defined by the pixel rows or columns of a CCD detector. 
It is not possible to merge them into a 1D spectrum by interpolating among adjacent orders, as this adds correlations  which affect the extracted spectrum.  The response of each order to the blaze function varies,
and the consequent variation of count levels along and across orders can be prohibitively large.
Also, if the resolution is high, some orders are very narrow, making it difficult to deblaze the spectra due to the presence of strong lines, such as the Balmer lines.  Unfortunately there is no standard or perfect way to merge orders.  An extensive discussion of these challenges can be  found in \cite{2001A&A...369.1048P}. 

\paragraph*{Removal of telluric features:}

Part of the stellar light passing through the Earth's  atmosphere will be absorbed, causing the so-called telluric lines in the spectrum. They can be very strong and have fixed positions in wavelength, but not all features are identified. Atlases of telluric standards or models exist \citep[see discussion in, e.g.,][]{2014A&A...564A..46B} 
and are cross-correlated with the science spectrum in order to identify and if possible, remove these lines \citep{2018PASP..130g4502S}. 
There are regions in the spectrum which are more affected by telluric absorption, notably towards the red part of the spectrum \citep[][their Fig. 16]{2017MNRAS.464.1259K}. 



\paragraph*{Normalisation:}
Since flux calibration is very challenging for high-resolution spectra it has become customary to determine stellar abundances from spectra normalised to the continuum flux, effectively using the relative strengths of the absorption lines. There is no established standard way to normalise a spectrum to the continuum, although procedures such as the {\tt continuum} task in the IRAF software system are very popular. In general, the pseudo-continuum is determined after fitting a spline or a polynomial to a set of regions that are believed to be free of absorption lines \citep[see, e.g.,][]{2001A&A...369.1048P}. 
We stress the term ``believed", since it is not certain that areas free absorption exist at all across a given spectrum. 
Examples where continuum normalisation is especially complicated are very cool stars, which have spectra crowded with molecular features, spectra with too low S/N, or spectra whose orders are not properly merged.  Other regions which are particularly challenging for FGK-type stars are around the Balmer lines, especially for high-resolution echelle spectra.    
The definition of the continuum may in fact be responsible for the largest fraction of the uncertainty in abundances \citep[e.g.,][]{GBS6}.

\subsection{Physical description}

Very nice summaries of various aspects of the physical description of line formation theory can be found in the reviews of \cite{2005ARA&A..43..481A}, \cite{2016LRSP...13....1A} and \cite{nissen-gustafsson}. 

\paragraph*{Line formation:}
A spectral line is usually studied with the help of a radiative transfer calculation and a model for the atmosphere. It is clear that 3D-non-LTE models are the way forward to obtain accurate absolute abundances. However, one should be aware that these models focus on improving certain aspects of atmospheric physics (geometry and statistical equilibrium), while other aspects, such as the treatment of opacities by sampling and binning, can still be quite uncertain (see below). 
A very interesting message from \cite{2005ARA&A..43..481A} 
is that the fact that LTE is the standard method of analysis does not mean that departures from LTE only occur occasionally. 
Yet, the simplistic 1D-LTE models are still the ones mostly used when abundances are derived industrially.

The reason might be that we now have a reasonable understanding of the applicability and failure of 1D-LTE, mostly thanks to the progress in 3D-non-LTE modelling enabling one to quantify the differences with respect to 1D-LTE models. Non-LTE corrections for abundances of hundreds of lines of several elements, as well as grids of 3D atmospheric models and synthetic spectra, are publicly available (see Table~\ref{tab:tools}). 
In 1D-LTE the parameters micro- and macroturbulence account for the turbulent motions of particles, and need to be specified for modelling the lines. They are not needed when a full hydrodynamical simulation of the atmosphere is performed. With 3D models it is possible to find empirical relations for  these parameters as a function of stellar parameters \citep{2013MSAIS..24...37S}.  Such resources are important, as they allow one to identify the conditions (and lines) for which the differences between 1D-LTE and 3D-non-LTE are minimal. Thus, spectral analyses can be calibrated such as to yield accurate results even under the assumption of 1D-LTE.  The great difficulty is that, for many elements, especially those which produce few lines,  no ``3D-non-LTE free lines'' are available. 
To achieve high precision and accuracy by taking advantage of all available lines in the spectrum, it is thus of paramount importance to work towards providing the prerequisites for modelling lines of more species in 3D and non-LTE.

\paragraph*{Geometry and opacities:}
In 1D models (and 3D stellar surface simulations of the ``box in a star'' type), the geometry can be plane-parallel or spherical for each layer of the atmosphere. Both models are of comparable accuracy for dwarfs, but for giants and supergiants, which have extended atmospheres,  curvature needs to be taken into account. Abundances have been compared for 1D models with different geometry for several elements and lines by \cite{2006A&A...452.1039H}, 
who found that strong lines of high excitation potential are most affected. 

Another central issue in modelling atmospheres is accounting for all possible opacity sources. They can be divided in continuous (produced by bound-free and free-free transitions) and line (produced by bound-bound transitions) opacities. The contributors are hydrogen atoms, metal atoms, and molecules. In cool stars, molecules are especially problematic, because they are poorly known  \citep[see][for details]{1992AA...256..551P, 2014AA...571A..47M}. 
\cite{2008AA...486..951G} present further discussion on this subject, where they test the structural effect on  MARCS atmosphere models for different temperatures and optical depths when including and excluding opacities due to H, metals, and molecules.

\begin{textbox}
\section{Analysis methods: EW or synthesis?}
Both types of methods are good competitors, and it is not clear which of them performs best. Reports on comparisons of these methods can be found within the Gaia-ESO framework \citep{2014A&A...570A.122S}, 
as well as in the series of works on the \gbs s \citep[e.g.,][]{GBS4}, 
by \cite{2016ApJS..226....4H}, 
or by \cite{2017MNRAS.470.4363C}. 
Using EWs or synthesis with high-resolution, high S/N spectra of solar-like stars often seems to be a decision of personal preference. Syntheses might have more applicability in crowded spectral regions, or in stars with broad lines. Today, computers are able to quickly synthesise spectra, so the computing time is not the limiting factor as it used to be a decade ago. It is possible that, in the next decades, syntheses will become the preferred way to measure abundances, but EWs should not be set aside completely, as they are the simplest tool to measure the strength, and hence understand the nature of the lines under analysis. 
\end{textbox}

\subsection{Analysis methods}
\label{sect:analysis_methods}
The classical and most common methods to determine abundances are based on the measurement of equivalent widths (EWs) or the computation of synthetic spectra of absorption lines of the chemical element in question. Recently, machine-learning approaches for measuring abundances have been introduced and applied to stellar surveys, and are further discussed in Sect.~\ref{sect:surveys}. 
As of today, EWs and syntheses are still the dominant methods to determine abundances, especially because machine-learning methods still rely on training sets of stars with ``well-known" abundances, which most likely will be measured or calibrated from EWs or synthesis methods. 

\paragraph*{Equivalent widths:}
They are obtained from either fitting a Gaussian profile for weak lines and Voigt profiles for stronger lines, or just by integrating over the line profile. The latter becomes more accurate when lines have a boxy-shaped profile due to, e.g., hyperfine structure components (see Sect.~\ref{sect:line_list}). The EW is thus a measure of the strength of the line, which can be directly related to the chemical abundance of the element in a star given the stellar parameters based on the so-called curve of growth (CoG): for weak lines there is a linear increase of abundance with EWs (in a logarithmic sense). Stronger lines lie on the flat part of the curve of growth: they are saturated and thus there is no direct relation of the abundance with the EW. Note that the definition of weak and strong lines might vary from star to star and therefore one has to select the lines which lie in the linear part of the CoG in each case individually.

The dominant source of uncertainty in EW methods is the placement of the continuum. Today, typical automatic codes are still not able to identify the continuum as precisely as can be done by hand using, e.g., the {\tt splot} task of IRAF, especially when the spectra are crowded with stellar features or artifacts due to data reduction. 
Experienced spectroscopists may be able to identify the continuum for such challenging lines `by eye", making this process rather more an ``art" than an objective task. 
Measurements by hand are usually limited to high precision abundances of small samples of stars \citep{2017A&A...608A.112N, 2018arXiv180202576B}. 
In this era of industrial stellar abundances, EWs are best measured with automatic pipelines.  The uncertainties are probably larger than for the manual measurements, but they can be reproduced and quantified, and so can their effect on derived abundances. 
A serious limitation of determining abundances from EWs is that, if lines are blended, the abundance will be overestimated. Thus, EWs work best for very high resolution and high S/N spectra. Likewise, intrinsic broadening of lines contributes to blending, and so EW methods work best for relatively warm stars with slow rotation. Most spectroscopic surveys are designed to obtain spectra of stars where these conditions are met, and in this case it will be safe to use the EW method.

\paragraph*{Synthesis:}
 The abundance of the element is varied until the best fit of a synthetic line profile with respect to the observation is found.  Syntheses can be computed on-the-fly for each line and star until the best fit is obtained. It is also possible to use pre-computed grids of synthetic spectra with varying abundances for different sets of stellar parameters \citep{2016AJ....151..144G}.  Syntheses on-the-fly have the advantage that they can be easily adapted to different spectra and lines. This freedom allows one also to identify stars with unusual chemical abundances. The disadvantage is that, when large samples of stars need to be analysed, the analysis can be very time-consuming. This might be especially inefficient when the stars are very similar to each other, like those targeted by spectroscopic surveys. \cite{2018arXiv180401530T} present a solution to overcome this problem by interpolating between models. 

Syntheses are the preferred method when spectra are crowded with absorption features, which is the case for cool stars. In addition, they are the only way to measure abundances from molecules \citep{2014AJ....147..136R} or from very blended lines, for example.  This is because the wavelength region to be fitted can be set to intervals of arbitrary size, and so abundances are not restricted to be measured from individual lines that have a ``well-behaved" shape. The disadvantage with respect to EWs is that they depend on the instrumental profile (e.g., the spectral resolution needs to be known) and every pixel is fitted, which means that the results are sensitive to an imperfect wavelength calibration \citep{2016ApJS..226....4H, GBS6}, for example.  

\subsection{Line list}
\label{sect:line_list}
When deriving abundances from absorption lines, it is assumed that the line strength is directly related to the abundance of the element whose transition produces the measured line. The wavelengths and transition probabilities, as well as the properties of the atomic states responsible for these transitions, are stored in a line list. 
The accuracy of the atomic parameters has become one of the major sources of uncertainty in the abundance determination.  Significant efforts are being dedicated by laboratory spectroscopists and theorists to provide the needed data for transitions of many elements and species. This is tedious and challenging work, exemplified by the fact that only about half of the lines in the optical wavelength range (480 to 680~nm) that are often used for abundance analysis of solar-type stars have good laboratory transition probabilities, that is, with typical uncertainties of 10 percent or better \citep{heiter-linelist}. Moreover, current lists of lines with good wavelengths contain only half of the lines observed in good quality solar spectra \citep{2014dapb.book...39K}. 
The situation becomes especially problematic at cool temperatures, where molecular lines dominate over atomic lines in the spectra.
The line data are less complete for wavelength ranges outside the optical, such as the UV and the IR.
Here we discuss a selection of issues related to the line list that are important when deriving abundances. 

\subsubsection{Transition data}
\label{sect:line_data}
One of the most fundamental information in the line list is the transition probability, often presented in the form of $gf$-values (product of statistical weight and oscillator strength). When these values are not known accurately, it is common to perform an ``astrophysical calibration": deriving the oscillator strength for a line by setting the abundance of an element to a reference value and fitting a synthetic to an observed spectrum by varying the $gf$-value. Usually this is done for the solar spectrum, for which the chemical composition is known with the highest accuracy.  \cite{2016A&A...587A...2B} 
present a detailed discussion on calibrating $gf$-values based on several \gbs s. From a comparison with accurate laboratory measurements, they conclude that the final calibrated values may be subject to systematic uncertainties caused by normalisation, line fitting procedures, 3D-non-LTE effects, errors in the stellar parameters and the solar abundances adopted. While using ``astrophysically calibrated" atomic data has been shown to improve the precision of stellar abundance results on several occasions, it is not obvious that these results are accurate.  Calibrating atomic data in this way offers a temporary solution until direct and accurate measurements in the laboratory become available for all lines in stellar spectra. 

Experimental and theoretical data for atomic and molecular transitions are made available through on-line collections and databases, such as those by R.L. Kurucz, at NIST, or the VALD database (cf. Table~\ref{tab:tools}). 
A major step towards standardized access and distribution of atomic data is done by the VAMDC Consortium (Virtual Atomic and Molecular Data Centre), which maintains an electronic infrastructure providing access to about 30 databases simultaneously, together with tools and policies that aim to enhance the citation rate of individual data producers.

These databases contain further data that are needed to calculate synthetic spectra, in particular parameters that describe line broadening (see \citealt{2016A&ARv..24....9B} for a recent review). 
Apart from the natural broadening due to the finite lifetimes of atomic states, the most important broadening process is collisions with neutral hydrogen, which can be described with different recipes. This includes the approximate formulation based on the van der Waals potential from the 1940s and 1950s (Uns\"old recipe), and the more detailed theory by Anstee, Barklem, and O’Mara from the 1990s (ABO theory, see \citealt{2016A&ARv..24....9B} and \citealt{heiter-linelist}).
An example of the effect on abundances of using the Uns\"old recipe versus the ABO theory is given by \citet{2007ApJ...667.1267S}. For 58 \cri\ lines with a mean EW of 40~m\AA, the change in the mean solar \cri\ abundance was 0.02~dex.

An additional complication arises from the presence of hyperfine structure (HFS) components in individual atomic lines for species with odd baryon numbers (non-zero nuclear spin; for Solar System isotopic abundances these correspond mostly to elements with odd atomic numbers).
The HFS parameters from which the exact positions of the components in wavelength can be calculated represent another type of atomic input data, while the relative intensities of the components are directly computed from quantum numbers.
%
When unresolved, HFS can be regarded as an additional broadening mechanism, changing both the shape of the line profile and the total line intensity.
The effect is larger for strong lines, since they may be de-saturated. 
There is extensive literature studying the effects of HFS on abundances, (see \citealt{2015A&A...577A...9B} and \citealt{GBS6} for some examples).

Similarly, for atoms with several stable isotopes, the different atomic masses split the energy levels and thus a given transition into several components with a different wavelength for each isotope.
In this case, the relative intensities of the components only depend on the isotopic composition under consideration. 
At Solar System composition, there is typically one dominating isotope for each element,  thus the effect is mostly negligible, with the notable exception of Cu (with about two thirds of $^{63}$Cu and one third of $^{65}$Cu).

Finally, line lists need to include transition data for molecules as well as atoms. For molecules we rely to a larger extent on theoretical calculations than for atoms, with correspondingly larger uncertainties in data quality.
In G- and K-type stars the transitions of diatomic molecules play an important role. For example, for the Gaia-ESO survey (Sect.~\ref{sect:GES}),  data for twelve different molecules of this kind are provided (27 isotopologues, mainly hydrides and carbon-bearing species). 
The main purpose of including molecular lines in the abundance analysis is to identify and account for blends affecting atomic lines. However, for some elements, in particular C, N, and O, molecular features are also used for abundance determination and to determine isotopic ratios.
\cite{2014AA...571A..47M} 
illustrate the effect of including transitions of CH in calculated spectra at wavelengths bluer than $\sim$450~nm, for the Sun and four metal-poor stars, showing a significant improvement when comparing to observed spectra.

\begin{figure}
\hspace{-2cm}
\includegraphics[scale=0.35]{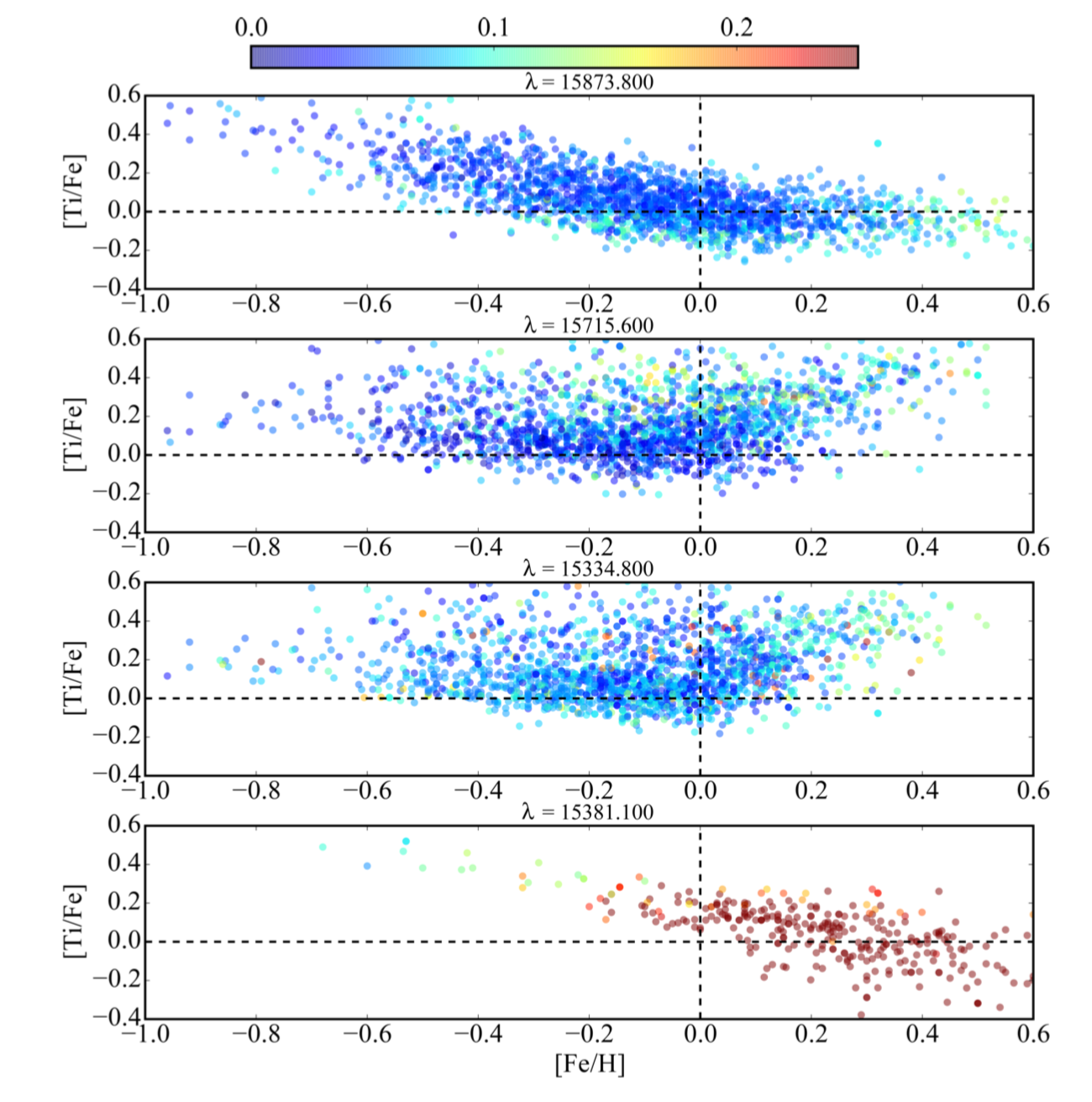}
\caption{[Ti/Fe] ratios as a function of metallicity determined from different Ti lines selected from APOGEE spectra. Colours indicate the scatter of 4 different methods used to derive abundances.  Each panel shows a different trend, showing that line selection plays a crucial role in the final abundances. Credit: Hawkins et al, A\&A, 594, 43, 2016, reproduced with permission \copyright\ ESO.}
\label{2.line_selection_hawkins}
\end{figure}

\subsubsection{Line selection}
\label{sect:line_selection}

Ideally, one should select lines that have a wide range in strength, and are spread out over the spectrum, i.e., at different wavelengths and excitation potentials. This helps to avoid systematic effects of any variations in spectral response and to probe different parts of the atmosphere. Furthermore, one should select lines at different ionisation stages, as these show different sensitivity to changes in atmospheric pressure. If the analysis is accurate, the abundances derived from every line should be consistent, allowing one to provide an average of the results obtained for each line as the final abundance.  In reality, in many cases few lines are available and an average might not be accurate.

In \cite{2016A&A...594A..43H}, for example,  
a comparison of titanium abundances from different lines in spectra from the APOGEE survey (Sect.~\ref{sect:APOGEE}) was made. The main result is illustrated  in Fig.~\ref{2.line_selection_hawkins}, where the [Ti/Fe] abundances as a function of \feh\ are shown for four different lines.
Note that one of them was not detected for a large portion of the stars. The colours in Fig.~\ref{2.line_selection_hawkins} show the scatter among the different methods that were employed to derive the abundances. 
Among the three lines which were detected in the bulk of the stars, only one shows the expected trend with \feh, similar to that of other $\alpha$-elements, while the trends of the other two lines are very different. As a possible explanation the authors mention non-LTE or saturation effects, as both lines are very strong.
The titanium abundances published in the APOGEE data releases are based on a different line list, with astrophysically calibrated atomic data, and a different analysis method (see, e.g., \citealt{2015ApJS..221...24S}, 
\citealt{2018arXiv180709773H}, 
and Sect.~5.12 in \citealt{2018arXiv180709784J}). 
Therefore, the findings by \cite{2016A&A...594A..43H} 
cannot be directly applied to assess the abundance data from the APOGEE survey. 

\begin{figure}
\includegraphics[scale=0.40]{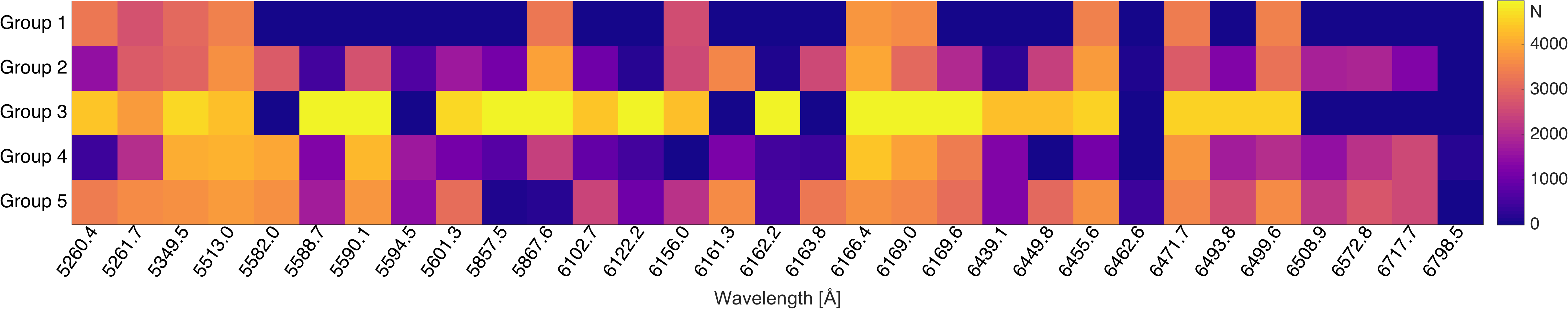}
\caption{\cai-line selection in the Gaia-ESO survey. Colour coding represents the number of stars for which an abundance was determined for each line by different analysis groups participating in the internal data release 5. Based on data provided by R.~Smiljanic (priv. comm.) reproduced with permission of Gaia-ESO.}
\label{fig:heatmap}
\end{figure}

Chromium abundances in metal-poor stars are also quite sensitive to line selection.
\cite{2017ApJS..228...10L} 
used $\sim$40 \cri\ lines and 75 \crii\ lines to derive the abundance of the metal-poor star HD~84937. The mean abundance of \cri\ was almost 0.1~dex lower and the dispersion about twice as large compared to \crii. The discrepancy decreased to less than 0.05~dex when the six \cri\ resonance lines were removed, with a corresponding improvement in dispersion (becoming similar to that of \crii).
The authors note that half of the \cri\ resonance lines (the triplet at $\lambda\sim$4275\AA) have often been employed in abundance studies of metal-poor stars.
The remaining discrepancy between \cri\ and \crii\ line abundances, mainly seen at wavelengths $>$4000~\AA, can be ascribed to non-LTE effects in \cri, as studied by \citet[who did not include the resonance lines]{2010A&A...522A...9B}. 

In addition to issues in either atomic data or physical assumptions for the line formation,  misidentification of the continuum, or unidentified blends may also affect the selected lines. For a given observing time, it is more difficult to obtain good S/N in the blue parts of the spectrum than in the redder parts. Furthermore, the blue region contains more absorption lines (hence blending is more severe for more metal-rich stars in that region). The wavelength coverage of selected lines thus might have a strong dependency on stellar type and metallicity.  

Even though the line selection may follow the reasoning discussed above, different criteria for ``problematic lines'' may be defined by different methodologies. 
An example for the variation of line selection is given in Fig.~\ref{fig:heatmap}, which visualises the line selection within the Gaia-ESO survey (Sect.~\ref{sect:GES}) for \cai. All groups performing the analysis were provided with the same set of observed spectra, and the same line list containing 31 \cai\ lines. Nevertheless, the number of stars for which each group determined an abundance for each line varied significantly. There are a few lines which were consistently used (e.g., 5513, 6166~\AA) or discarded (e.g., 6463, 6799~\AA) by all groups, while others were employed by only a sub-set of the groups (e.g., 5260, 5582~\AA).

\subsection{Stellar parameters}
\label{sect:params}
The relations of line strength and abundance depend on the stellar parameters.   A natural approach would be to determine the parameters ``consistently" with the abundances, e.g., from the same spectra, line lists, prescription, method, etc. However, this is not necessary as in some cases accurate parameters can be determined with methods that are independent from spectroscopy.  
The PASTEL catalogue \citep{2016A&A...591A.118S} 
is a valuable resource to learn about the complexity of stellar parameter determination.   The catalogue contains a collection of more than 1000 bibliographic resources of reported stellar parameters of more than 30\,000 stars,  determined from any of the methods discussed below, and shows the inhomogeneity of stellar parameters resulting from different studies.  

An investigation of the values in the PASTEL catalogue shows that differences of 200--300~K in effective temperature (\teff)  are usual for FGK-type stars analysed by different methods.  For the stars in PASTEL with more than 25 \teff\ determinations, a typical difference of 50~K is obtained.   This suggests that it is today not possible to know the temperature of a star better than this accuracy.  Large efforts are invested in obtaining more accurate temperatures of stars, because of the variety of astrophysical applications which depend on a temperature scale, however, no conclusion has yet been reached as to which method should be employed. Methods which are often used are the infrared flux method, excitation balance, fitting of Balmer lines and interferometry, which are explained in detail in the supplementary material (Sect~\ref{Supl1}).

The effect of surface gravity (hereafter \logg) on the spectra is weaker compared to \teff, which poses a challenge to constraining this parameter spectroscopically. It is difficult to determine \logg\ to better than 0.1~dex in FGK-type stars. In the PASTEL catalogue 
the typical reported errors in the literature are of that size, which agrees with the median difference in \logg\ obtained from independent works on the same stars. The comparison between APOGEE and LAMOST of \cite{2018arXiv180707625A} 
shows that \logg\ has a scatter of 0.25~dex among these surveys.  Common methods to derive \logg\ are the parallax method, ionisation balance, fits of strong lines, and asteroseismology. These methods are also explained in  the supplementary material (Sect~\ref{Supl1}).

Metals influence both the strength of spectral lines and the continuous opacities, in cool stars mainly through the abundance of H$^-$, which depends on the presence of metallic electron donors. A change in metallicity changes the overall atmospheric structure, which is why metallicity is one of the main stellar atmospheric parameters. Unlike \teff\ and \logg, metallicity can only be measured directly from the analysis of a spectrum. Indirect determinations based on theoretically or empirically calibrated photometry, have also been widely used when no spectrum is available. We note that they are affected by the same issues than photometric temperatures (see supplementary material, Sect~\ref{Supl1}).  Metallicity is commonly referred to as \feh, because one of the main techniques to estimate this parameter is to determine iron abundances. However, in general, the abundances of other elements may not scale with Fe, which makes the designation of metallicity by \feh\ imprecise. The stellar metallicity can also be expressed as [M/H], usually representing a combination of \feh\ and \afe. Whether metallicity refers to \feh\ or \mh\ depends mostly on the method employed to determine this parameter, and what assumption is used for the enhancement of $\alpha$-elements of a given star.  Like stellar abundances in general, to determine metallicities one must take care of all the steps and issues discussed in this section. There are two main ways to derive metallicities, either by measuring iron abundances from iron lines, or by performing a global fitting to the spectra. These methods are discussed with more detail in  the supplementary material (Sect~\ref{Supl1}).

\paragraph*{Other parameters:}
\label{sect:otherparams}
To relate \feh\ and \mh\ it is assumed that, at
solar metallicities, $\alpha$-elements are solar-scaled  and the $\alpha$-element abundance linearly increases towards lower metallicities,
reaching a plateau of $[\alpha/\mathrm{Fe}] = +0.4$ at [Fe/H]$=-1$. However, at lower metallicities, variations in C and N might further affect  the opacities, and a proper atmosphere model should be adopted to avoid additional uncertainties in abundances \citep{2017ApJ...847..142E}. 

There are line broadening parameters that affect the overall structure of the atmospheres. In 1D modelling, the most notable one is the microturbulence (\vmic). It accounts for the small-scale turbulent motions of the particles that lead to excess line broadening. The stronger the line, the larger the effect due to \vmic\ (see Figure~\ref{3.sensitivity_lines} and discussion in Sect.~\ref{sect:unc_param}).
In 1D spectral synthesis calculations \vmic\ does not have a physical meaning, but is an ad hoc parameter needed to improve the line shape.  Hence, the value of \vmic\ can be slightly different for different methods even when \teff, \logg, and \feh\ agree.  Microturbulence is normally derived by requiring that iron abundances remain the same regardless of the strength of the line. When not enough lines are available, it is possible to use empirical relations that depend on the other stellar parameters. In fact, this is done for most of the surveys (see Sect.~\ref{sect:surveys}).  In a 1D analysis, \vmic\ counts as a fourth stellar parameter. The value adopted for \vmic\ influences the abundances, so it is important to state which value was considered when abundances are reported. 

Further broadening parameters that need to be specified when synthesising spectral lines are the projected rotational velocity ($v \sin{i}$) and the macroturbulence (\vmac). Similarly to \vmic, \vmac\ tries to account for large-scale turbulent motions in the atmospheres, which in 3D modelling are fully incorporated. \cite{2008AJ....135..892C} 
present a study of these effects in metal-poor giants.   Since \vmac\ and $v \sin{i}$ have a very similar broadening effect, it is difficult to disentangle both effects directly from the spectra. \cite{2008AJ....135..892C} performed a Fourier transformation on high-resolution and high-S/N spectra to determine both parameters. However, such analyses are rarely done, rather, it is common to set either \vmac\ or $v \sin{i}$ to zero and determine a global broadening parameter, or to use a value of \vmac\ based on empirical relations like those for  \vmic. This is especially the case when spectral resolution or S/N are not sufficient to disentangle the effects from the two broadening mechanisms.

%% file: table_tools.tex
\begin{table}[t]
\hspace{-2cm}
\caption{Material and tools for spectral analyses that are publicly available}
\label{tab:tools}
\begin{tabular}{@{}lcc @{}}
\hline
Material & Reference & Comment \\
\hline
& \bf{Spectral libraries} & \\
SVO & \url{http://svo2.cab.inta-csic.es/theory/libtest/index.php} & Public libraries \\
Montes & \url{https://webs.ucm.es/info/Astrof/invest/actividad/spectra.html} & Compilation \\
\hline
& {\bf Model atmospheres} & \\
MARCS & \cite{2008AA...486..951G} & 1D spherical geometry \\
ATLAS9 & \cite{2003IAUS..210P.A20C} & 1D plane-parallel geometry \\
{\tt STAGGER} & \cite{2013AA...557A..26M} & 3D \\
CO5BOLD & \citet{2012JCoPh.231..919F} & 3D \\
\hline 
& {\bf Radiative transfer codes} & \\
Turbospectrum & \cite{1992AA...256..551P} & LTE\\
MOOG & \cite{1973PhDT.......180S} & LTE \\
SYNTHE  & \cite{1993sssp.book.....K} & LTE\\
SPECTRUM & \cite{1994AJ....107..742G} & LTE \\
DETAIL/SIU & e.g., \citet{2012ApJ...751..156B} via \url{http://nlte.mpia.de/} & Non-LTE\\
\hline 
& {\bf Line lists} \\
VALD & \cite{2015PhyS...90e4005R} & Literature compilation \\
NIST ASD & \url{https://www.nist.gov/pml/atomic-spectra-database} & Literature compilation \\
Sneden et al. & \url{https://www.as.utexas.edu/\~chris/lab.html} & Bibliography and molecular line lists \\
ExoMol & \cite{2016JMoSp.327...73T} & Very cool objects \\
BRASS & \cite{2018Galax678L} & Centralisation of sources\\
Barklem & \cite{2015ascl.soft07007B} & Broadening cross-sections \\
Kurucz & \citet{2011CaJPh..89..417K} & Atomic data\\
VAMDC & \url{http://www.vamdc.eu} & Electronic infrastructure \\
\hline
& {\bf Grids of synthetic spectra} &\\
AMBRE & \cite{2012AA...544A.126D} & Optical high resolution\\
{\tt STAGGER} & \cite{2018AA...611A..11C} & \caii\ triplet centred\\
 3D-non-LTE Balmer & \cite{2018AA...615A.139A} & Balmer lines centred \\
 APOGEE & \cite{2012AJ....144..120M} & Infrared \\
 POLLUX & \citet{2010AA...516A..13P} & Database \\
 \hline
 & {\bf Automatic codes for the determination of abundances}\\
 SME & \cite{2017AA...597A..16P} & With non-LTE on the fly \\
 {\tt iSpec} & \cite{2014AA...569A.111B} & Python wrapper for various tools \\
FERRE & \citet{2016AJ....151..144G} & Match models to data \\
GALA & \cite{2013ApJ...766...78M} &  EW code \\
DOOp & \cite{2014AA...562A..10C} & Wrapper for EWs \\
ARES & \cite{2015AA...577A..67S} & Automatic EWs \\
The Cannon & \cite{2015ApJ...808...16N} & Label transfer from a training set\\
\hline
& {\bf Non-LTE abundance corrections} \\
INSPECT & \url{http://inspect-stars.com/} & Line-by-line corrections \\
MPIA & \url{http://nlte.mpia.de/} &  Line-by-line corrections\\ 
\hline
\end{tabular}
\begin{tabnote}
This compilation is possibly not complete. It is restricted to tools that are regularly updated and available on the web. Other codes not listed here might be equally suitable and available upon request to their authors. 
\end{tabnote}
\end{table}

%% file: uncertainties.tex

In recent years, large datasets of seemingly homogeneous stellar abundances have appeared on the scene, notably from spectroscopic surveys, moving the production of abundances towards industrial scales.
For each dataset, the combined effects of the steps discussed in Sect.~\ref{sect:methods} on the measurements of abundances of a given star could be interpreted as the ultimate uncertainty.
Extensive discussions of such uncertainties can be found in the literature compilations of \cite{2008PASJ...60.1159S} 
and \cite{2014AJ....148...54H}.  
Thus, it becomes increasingly challenging to obtain homogeneous abundances that can be used for a large variety of science cases.
Furthermore, combining the abundances from different surveys is none-trivial, often due to  correlated uncertainties arising from each step of the abundance analysis procedure.
Some of the uncertainties may be amplified when combining results from different groups that employ different data and methods \citep[see, e.g.,][]{2014A&A...570A.122S}. 
As an additional complication, uncertainties are assessed in different  ways by different works. It is thus desirable that different catalogues perform similar tests to assess  uncertainties,  enabling better comparison and combinations.  
\cite{2014AJ....147..136R} 
is an inspiring work, in which several sources of uncertainties are extensively discussed. The series of works on the \gbs s \citep[e.g.,][]{GBS4} also provide detailed discussions of the matter.   Here we disentangle and briefly discuss different parts of the abundance error budget, dividing the uncertainties into three main categories: random,  systematic, and biases. 

\subsection{Random uncertainties}

Here, we refer to random uncertainties as uncertainties related to the input material (characteristics of input spectra, uncertainties in laboratory data, data reduction issues, and so on). In order to quantify these, there are some tests that can be performed. 

\subsubsection{Instrumental error} 
Using different spectra for the same stars allows one to quantify uncertainties due to the characteristics of the input spectra (S/N, resolution, normalisation, instrumental responses in general). The abundance analysis method may be tested using a set of reference stars for which spectra exist in several archives. For example, \cite{2014AJ....147..136R} 
compared EWs from different instruments, and they found that the largest deviations arose for strong lines and low S/N, for which blends could not be identified. However, they demonstrated that, for the typical S/N of their sample, weak lines gave consistent results for different instruments. 
Another possibility is to use repeated observations of the same star at different S/N. \cite{2016A&A...591A..34A} 
discuss how abundances are affected when spectra from the same instrument but of different S/N are used. They found an increased significance of abundance trends of \xfe\ versus condensation temperature for higher S/N spectra.  This implies that, before interpreting such slopes astrophysically (e.g., presence of debris disks or planets), one must carefully assess the instrumental dependencies of the abundances obtained. Such statistical uncertainties are particularly important for high-precision studies. For planning spectroscopic surveys a key issue is to find the threshold in S/N required for achieving the desired abundance precision for a given set of spectra and the methodology to be used. This uncertainty will dictate the size of the dataset and the Galactic region sampled.

An alternative way to quantify uncertainties due to input spectra is to look at the differences obtained in abundances for cluster members, which are expected to have the same abundance pattern \citep[although see, e.g,][]{2016MNRAS.457.3934L}. Different stars of the same spectral class essentially should yield the same abundances. Thus, the variation can be attributed to statistical uncertainties. These tests are performed by some surveys (see Table~\ref{tab:errors} and Sect.~\ref{sect:surveys}). 
Errors are commonly given as the standard deviation about the mean of the abundances obtained from all measurements. 

\subsubsection{Uncertainties due to line selection}
\label{sect:unc_line}
In general, one can assume that the results will be more accurate the more lines are used for a given element. However, including too many lines might have negative consequences on the results if a considerable number of lines are saturated, too weak, blended, have poor  atomic data, poor HFS treatment, poor spectra, are contaminated, etc. A classical way to quantify this uncertainty is providing a line-to-line dispersion (LLD). Uncertainties derived from neutral lines are often observed to be smaller than those from ionised lines, but this may be due to the fact that, in FGK-type stars, more neutral than ionised lines are available for estimating this dispersion. While this uncertainty is commonly reported, the definition for LLD differs from work to work. It is common, for example, to decrease the uncertainty by adopting a ``$\sigma$-clipping'' procedure, that is, removing outlier lines whose abundances differ by more than a given value from the mean abundance obtained from all lines. That value can be a factor of $\sigma$, with $\sigma$ representing the standard deviation of the Gaussian distribution of all abundances. The factor varies in the literature, for example \cite{2018AJ....155..111L} 
performs a cut at 2.5$\sigma$, while \cite{2011A&A...534A..53P} 
use 3$\sigma$. 
In some cases \citep{2012A&A...545A..32A, 2013ApJ...766...78M, 2017A&A...600A..22M}, the random uncertainty is reported to be the standard error of the mean ($\sigma/\sqrt{\mathrm{N}}$, i.e., dividing the LLD by the number of lines employed). This definition of error is obviously much smaller than the LLD, making the two uncertainty estimators incomparable. In many cases, few lines are available per element, and then $\sigma$ is very affected by a single outlier. The median is a more robust estimator of the final abundance, with the interquartile ($q_{75}-q_{25}$) range as its uncertainty \citep[for a good discussion see Chapter 3 of][]{2014sdmm.book.....I}. \cite{1990AJ....100...32B} also provide a number of robust and resistant estimators of location and scale that should prove useful.

\subsection{Systematic uncertainties}
We refer to systematic uncertainties as those uncertainties that arise from the approach employed to determine abundances, namely the method and line prescription assumed, which might induce different uncertainties in different parts of the parameter space. 

\subsubsection{Theory: 1D-LTE effects}
A certain level of uncertainty in the final abundances is caused by approximations in the line-profile prescription. Transitions are affected to various degrees by the assumption of 1D-LTE, which can be quantified and even corrected. The magnitude of these corrections varies across stellar-parameter space.  For many elements, model atoms required for non-LTE calculations are available (see Table~\ref{tab:tools}).  With the corrections at hand, the difference in the final abundance when using LTE and non-LTE results can be evaluated. 
Quantifying 3D effects is still difficult, since large grids with corrections for lines and elements are not available. However, the following diagnostics can be performed to assess the level of accuracy of the employed line-modelling prescription. 

If abundances can be derived from both neutral and ionised lines for the same element, the difference in these results may be attributed to uncertainties in the line-formation calculations. However, this method is not applicable if the stellar parameters have been determined by forcing ionisation and excitation balance, as this  causes an artificial agreement between abundances derived from neutral and ionised lines. To quantify this uncertainty, it would be ideal to determine \teff\ and \logg\ from methods which are less sensitive to 1D-LTE prescriptions (see examples in Sect.~\ref{sect:params}). Examples of detailed investigations of this kind for elemental abundances have been published in \cite{2016ApJ...817...53S} 
for iron-peak elements and \cite{2017ApJ...847...15B} 
for magnesium.  These works show that, although 1D-LTE modelling can be very uncertain, leading to incorrect measurements of abundances,  with a careful selection of lines, it is possible to derive accurate abundances. Careful selection of lines would favour ionised lines for which LTE holds better, and high excitation potential lines, for which 1D modelling is more accurate. This of course depends on the metallicity and overall atmosphere structure. In general, metal-poor stars are most affected. 

A selection of accurate 1D-LTE lines might require removing a large variety, if not all,  lines in optical spectra (for the case of metal-poor dwarfs, see \citealt{2016ApJ...817...53S} and \citealt{2018arXiv180507390R}). 
It is thus crucial to have a very extended wavelength coverage, including the infrared to the ultraviolet regions, and high resolution, in order to include as many ``clean'' lines as possible.
Unfortunately, outside the optical window, 3D-non-LTE effects have been investigated for very few elements and lines. Examples are \citet[][and references therein]{2017ApJ...847...15B}, 
who investigated the effect of 3D-non-LTE line formation of optical and IR Mg, Si, and Ti lines. Other examples are \cite{2017ApJ...835...90Z}, 
who looked at Mg lines in the H band to quantify this uncertainty in APOGEE stars, and \cite{2017A&A...607A..75N}, 
who quantified the uncertainties due to 3D-non-LTE of Al for a variety of stars and lines in the optical and IR. 
Regarding UV spectra, we must rely on observations obtained with HST, which are competitive and thus limited data are available. In any case, the majority of stars targeted by surveys have high metallicities and are rather cool. Thus, their UV spectra are so crowded with absorption features that almost no unblended lines can be used \citep{2016ApJ...817...53S}. 
To decrease ionisation-imbalance uncertainties due to poor modelling, it is recommended to follow the advice of \cite{2014AJ....147..136R}: 
to use the same ionisation stages for abundance ratios. If \fei\ results are to be considered for \xfe, then using the results for other elements from neutral lines will yield more accurate abundance ratios. The same applies for ions. 

\begin{figure}
\vspace{0.5cm}
\includegraphics[scale=0.45]{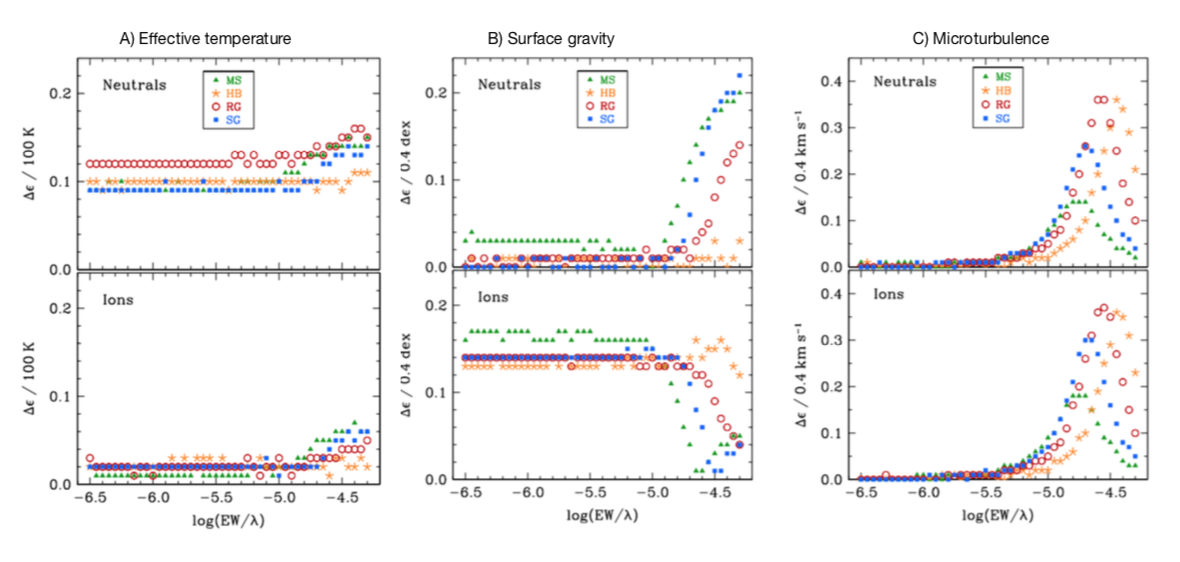}
\caption{Sensitivity of abundances obtained for neutral and ionised lines on stellar parameters, for metal-poor stars, as a function of reduced equivalent width. Different symbols represent different spectral types, namely MS for main sequence, HB for horizontal branch, RG for red giant and SG for subgiant stars. Panel A shows the abundance difference when the model temperature is changed by 100~K. Panel B illustrates the effect when changing \logg~by 0.4~dex and panel C shows the effect when changing \vmic\ by 0.4~km/s. Credit: Roederer, AJ, 147, 136, 2014. Reproduced with permission of first author and \copyright\ AAS.} 
\label{3.sensitivity_lines}
\end{figure}

\subsubsection{Uncertainties due to stellar parameters}
\label{sect:unc_param}
The final abundances depend to a large degree on the scale used for the stellar parameters.
In Figure \ref{3.sensitivity_lines} we show the effect on abundances when varying stellar parameters for neutral (top panels) and ionised (bottom panels) lines of several elements. This study was conducted by \cite{2014AJ....147..136R} 
on metal-poor stars of representative spectral types, namely main sequence (MS), horizontal branch (HB), red giant (RG) and subgiant (SG) stars. The figure compares the variation of abundances when changing the atmospheric parameters as a function of line strength. It nicely illustrates that abundances obtained from strong lines are more affected by uncertainties in stellar parameters than from weak lines.
It is also seen that, by changing \teff\ by 100~K (Panel A), abundances obtained from neutral lines are affected by 0.1~dex or more, while ionised lines change very little except for the strongest lines. The opposite is seen for \logg\ (Panel B). When changing the surface gravity in the model, weak ionised lines are more affected than neutral lines, while the situation is reversing at the strong-line end. This opposite behaviour forms the basis of determining stellar parameters from the combination of ionisation and excitation balance. Finally, Panel C shows that the abundances of strong lines are strongly affected by the adopted value of \vmic. 

In automatic analyses, it is relatively straightforward to compute abundances using different stellar parameters as input. The error due to stellar parameters can thus be estimated by comparing the difference in abundances obtained when the input parameters are varied according to their uncertainties. In this case, independent errors can be estimated for each parameter, which can be combined as explained in Sect.~\ref{sect:combination}.  Alternatively, star cluster members with stellar parameter differences of the order of the errors can be used to estimate the differences in abundances due to uncertainties in stellar parameters, since the abundances should be the same for all cluster stars \citep[although see, e.g., ]{2016MNRAS.457.3934L}. In this case, one obtains a single uncertainty accounting for all parameters together. 

\subsubsection{Using different methods}
A comparison of the results obtained from different methods allows one to study the dependency of abundances on the code employed. For example, \cite{2017MNRAS.470.4363C} 
perform a systematic study comparing stellar parameters using the EW wrapper GALA with the synthesis wrapper {\tt iSpec} (see Table~\ref{tab:tools}). They use this procedure to show that their conclusions are not affected by the methodology employed in the analysis.  At a more industrial scale, most of the error budget in the Gaia-ESO survey is assessed from the method-to-method dispersion  \citep[MMD, ][]{2014A&A...570A.122S}, and the same holds for the abundance analysis of the  \gbs s \citep{GBS4}. 
In fact, the strikingly large MMD seen in the Gaia-ESO survey has motivated the next generation of spectroscopic surveys to rely on one pipeline only \citep{2016AN....337..837A}. By excluding this type of uncertainty from the total error budget it is evident that the results will become more precise, but it will not be possible to investigate the dependency on the methodology employed and to truly assess the accuracy of the results. If many methods are used, a dispersion of the results can be calculated.
Similar to the discussion in Sect.~\ref{sect:unc_line}, we advocate employment of the median and the interquartile range (or the estimates described by \citealt{1990AJ....100...32B}), to quantify the dispersion rather than the mean and standard deviation. If only two methods are used, the error can simply be the difference between the two results.

\subsection{Reference objects and biases}
It is important to investigate any overall biases in the results, as this is key to combine different datasets. For this purpose, the results in different parts of the parameter space are compared with external sources.

\subsubsection{Trends of abundances with stellar parameters}
Star clusters are good laboratories for assessing if there are systematic uncertainties of the method for stars with different stellar parameters (e.g., dwarfs vs giants). Trends in abundances found as a function of \teff\ or \logg\ can be attributed to a systematic uncertainty of the method. In fact, for any stellar sample the behaviour of the abundances as a function of stellar parameters should be investigated. If the method is robust for a large range of spectral types, and if no effects of stellar evolution are to be expected, then no correlations should be found. If a clear correlation is found, it can be interpreted as a systematic uncertainty. Several works have chosen to apply a correction for such systematics, at least for \teff, by finding an empirical relation which is then used to scale the abundances according to their \teff\ (\citealt{2005ApJS..159..141V}, their Sect.~6.4 and Fig.~10; \citealt{2012A&A...545A..32A}, their Sect.~3.2 and Fig.~4). 

It is not simple to explain or correct the trends, as they can be caused by a variety of reasons. In \cite{2014AJ....147..136R} 
several such reasons are discussed in detail. In short, if the temperature decreases or the metallicity increases, lines become more affected by blends, which often are not identified.  Lines also become  stronger and start to saturate, which means that the selection of lines may vary across the parameter space. Thus, systematic differences may simply be the result of a line selection effect, instead of being due to variations in stellar parameters.  As strongly recommended by \citet{2014AJ....147..136R} 
and clearly demonstrated by, e.g., \citet{2015A&A...579A..52N}, 
if spurious \xfe\ trends exist as a function of stellar parameters, selecting stars from within a small region in parameter space for chemical evolution studies is the most secure way to proceed.

\input{reference_stars}

\subsection{Improving precision}
A homogeneous analysis with significantly reduced uncertainties might be achieved using a single pipeline, either by performing a differential analysis or by applying a data-driven approach. 

\subsubsection{Differential analyses}
Differential analyses consist in determining abundances in the same fashion for a given star and a reference star. The highest possible precision is achieved if the reference star is similar to the star of interest, because the overlap of suitable lines will be maximised. This reduces the LLD significantly, because uncertainties due to blends, poor atomic data, non-LTE effects, etc. are cancelled out to a certain degree. Furthermore, the continuum-normalised spectra are expected to be similar for similar kinds of stars, thereby reducing systematic uncertainties due to the methodology or due to stellar parameters. \cite{nissen-gustafsson} provide a complementary review focused on high-precision spectroscopic studies based on the differential technique.  

The accuracy of differential abundances fully relies on the abundance accuracy of the reference star. 
Differential analyses are thus very popular for solar twins \citep[e.g.,][]{2014ApJ...790L..25T, 2015A&A...579A..52N, 2018arXiv180202576B} because the Sun is our most accurate reference star (see Sect.~\ref{sect:gbs}). Precisions achieved are so high (better than 0.01~dex) that only with such an approach it is possible to study, e.g., the effect of planet formation  \citep[see Sect.~\ref{sect:science} and][for science applications]{nissen-gustafsson}. However, differential analyses of stars too different from the Sun require another reference star, since the more different the stars are, the less lines in common are available. Extensive discussions on this matter can be found in \cite{GBS4}.
In giants, \cite{2016A&A...594A..43H} 
improved the precision of the abundances by performing a differential analysis with respect to Arcturus. In metal-poor stars, \cite{2016A&A...586A..67R} 
performed a high-precision abundance study using as a reference G64-12. In clusters, precision can be improved by using one cluster member as a reference and deriving abundances for the other stars at the same location in the color-magnitude diagram differentially \citep[e.g.,][for the Hyades cluster]{2016MNRAS.457.3934L}. 

\subsubsection{Data-driven approaches}
Recently, new revolutionary ways to derive abundances with machine-learning tools have become very popular for the analysis of large datasets of spectra \citep{2018PASA...35....3N, 2018arXiv180401530T, 2018arXiv180804428L}. Empirical models or neural networks are built, where a relation between the spectrum and certain labels (abundances) is trained on a previously analysed subset of spectra. These relations are then applied to large samples of stars, resulting in impressively precise abundances even from data of seemingly rather low quality. Machine-learning methods have been very efficient in transferring the known information from the so-called training sets to entire datasets. However, it is not fully explored to what extent such methods are able to identify outliers.
As in the case of differential studies, the accuracy of the labels obtained with data-driven methods fully relies on the training (reference) sample. 

\input{uncertainties_cats}

\subsection{Combination of uncertainties}
\label{sect:combination}
Table~\ref{tab:errors} lists the different surveys and catalogues described in Sect.~\ref{sect:elements}, where we summarise which of the uncertainty assessments discussed here are performed. The abundance tests are separated according to assessing random and systematic uncertainties, and biases. We can see that all catalogues carry out at least one test in each of the categories, although they are not always the same.  In the listed works, the tests performed might not necessarily be included in the final error budget. This makes the comparison between catalogues, including uncertainties, difficult. Here we intend to provide guidance as to how the different uncertainties can be combined and standardised to provide for a more straightforward comparison in future catalogues. 

While an assessment of accuracy is provided by the external uncertainty (e.g., overall agreement with reference stars), a conservative measurement of precision for abundance determinations should take into account both the random and systematic uncertainties. According to our list, this means combining five different sources of uncertainties.  Following standard formulas for error propagation, the total error budget can be obtained considering the variances and covariances of the uncertainties. Let $\sigma_\mathrm{I}$, $\sigma_\mathrm{L}$, $\sigma_\mathrm{T}$, $\sigma_\mathrm{P}$, and $\sigma_\mathrm{M}$ be the uncertainty of Instrument, Lines, Theory, Parameters, and Methods, respectively (see Table~\ref{tab:errors}). These may be considered to be independent from each other, which means that the total error budget can be obtained from adding their variances ($\sigma_i^2$), where $i$ represents each of the five above sources. 

Determining $\sigma_\mathrm{P}^2$ can be more complicated, since it might originate from the analysis of the response of abundances to changes of the different stellar parameters separately \citep[see, e.g.,][]{GBS4}. 
The appendix of \citet{1995AJ....109.2757M} 
provides a well-structured presentation of a procedure based on a standard formalism for propagation of errors, showing how to obtain final uncertainties based on line-by-line abundance measurements, and how uncertainties in stellar parameters (which are not independent from each other) affect the final results. We discuss a few important conclusions from that work.
Firstly, because the uncertainties are correlated, the covariances between uncertainties can be calculated for a few representative stars in the sample and applied to the entire dataset. \cite{1995AJ....109.2757M} provides the atmospheric parameter variances and covariances for  a metal-poor star based on an analysis of optical lines. It would be useful to have such covariances for other types of stars in order to aid in the homogeneous presentation of abundances and their uncertainties by catalogues.
Secondly, it is shown that increasing the number of lines might reduce the random component of the uncertainty, while the systematic component remains constant. This implies that one should not consider $\sigma/\sqrt{\mathrm{N}}$, where $\mathrm{N}$ is the number of lines used, as an estimate of the total uncertainty of an average abundance.
Thirdly, for estimating abundance-ratio uncertainties one needs to keep in mind that the adopted atmospheric parameters (and associated uncertainties) are the same for both elements involved in the ratio, and that for some element lines the response of the final abundance to the stellar parameter uncertainty will also be very similar, leading to a partial cancellation of the systematic uncertainty. It is thus necessary to compute the covariances between the element abundances if the  abundance ratio uncertainties are to be estimated realistically.
\citet[][Appendix B]{2005A&A...439..129B} 
describes a modified version of the formalism of \citet{1995AJ....109.2757M} 
applicable to methods performing a global spectrum fit rather than determining line-by-line abundances.

%% file: reference_stars.tex
\subsubsection{Towards an absolute scale for abundances  using reference stars}
\label{sect:gbs}

A comparison of results with external sources helps to quantify the overall error budget, and to understand for what kind of stars the method is most accurate.   
The catalogues presented in Sect.~\ref{sect:catalogues} are widely used for comparison as they are large, enhancing the chance to have a sufficient overlap between datasets, and to study differences in a statistically significant way.  A standard for reference objects provides a more straightforward link between catalogues. 
Such reference objects can be either individual stars with well-defined properties or fields of stars with high-quality data available for a large number of stars, such as clusters or asteroseismic fields. 

In terms of stars with well-defined properties the Sun is undoubtedly \emph{the} reference star, considered as the standard reference for cosmic abundances. However, the determination of solar abundances is problematic, since different methods lead to different results. A review on the chemical composition of the Sun is given by \cite{2009ARA&A..47..481A}, who also provide recommended solar abundances determined with 3D hydrodynamical atmospheres. These revised abundances are, for some elements, in particular, light elements, significantly lower than those obtained with conventional methods \citep[e.g., the widely used scale of ][]{1998SSRv...85..161G}. Although the Sun is not observable in the same way as other stars, it is the most used star for differential studies, often using the reflection of sunlight from a Solar System body. Abundances tell us whether a given star with solar atmospheric parameters has exactly the same chemical composition as the Sun. The literature is very rich in studies looking for the closest solar twin. Several dozens of stars were claimed to be solar twins based on their atmospheric parameters, but when their detailed chemical composition is considered, the similarity is less obvious. For instance \cite{2016A&A...589A..17Y} performed a high-precision analysis that confirmed HIP~100963 to be a good solar twin, but with abundances of the $s-$ and $r-$process elements, as well as Li, slightly enhanced relative to the Sun. Other solar twins studied at high precision, and with a chemical pattern very similar to that of the Sun, include Kepler-11 \citep{2017ApJ...839...94B}, 
HIP~76114 \citep{2016A&A...587A.131M}, 
M67-1194 \citep{2016MNRAS.463..696L}, 
and HIP~114328 \citep{2014A&A...567L...3M}, 
which are good options to use as a reference star instead of the Sun. Abundances of solar-like stars relative to the Sun can be different due to several factors, such as Galactic chemical evolution, age, or the relative effects of non-LTE on stars with similar, but not exactly the same stellar parameters. 

\paragraph*{Gaia benchmark stars: beyond the Sun}
For stars that differ significantly from the Sun, it is not possible to measure abundances differentially to it. For that reason, the sample of \gbs s was built in order to establish a system of reference stars covering a larger range of atmospheric parameters \citep{GBS1}. The sample was designed to provide an anchor to the {\it Gaia} astrophysical parameter inference system that will estimate atmospheric parameters of one billion stars \citep[Apsis,][]{2013A&A...559A..74B}. These stars are fundamental calibrators because their effective temperature and surface gravity can be deduced directly from the accurate knowledge of their radius and flux distribution (see Sect.~\ref{sect:params}). Determination of their metallicity and abundances is described in \cite{GBS3} using a library of high-quality spectra \citep{GBS2}. Some surveys already use the \gbs s for their calibration, but the sample is still too small (around thirty stars), the stars are too bright, and they suffer from a deficiency of metal-poor stars.  Substantial efforts have been dedicated to extending the sample towards fainter and more metal-poor stars  \citep{GBS5}. Updated information and stellar parameters are provided via the CDS
\citep{2018RNAAS...2c.152J}.

Several surveys use other stars with well-defined properties that can be found in large catalogues such as PASTEL \citep{2016A&A...591A.118S} or Hypatia \citep{2014AJ....148...54H}. Both are bibliographical catalogues, making it possible to find well-studied stars that have been analysed independently by different groups who found consistent results.
This approach may be a way forward towards establishing a common set of reference stars. However, it is important to agree on a common set of procedures and criteria when selecting stars from  such catalogues.

Open and globular clusters are convenient reference objects due to their large number of members sharing in principle the same age and chemical composition. Some clusters, such as M67, have been extensively studied with high resolution spectroscopy that is available in public archives. Measuring the dispersion of abundances of cluster members obtained by an automatic pipeline is a good way to evaluate the internal precision over a range of stellar parameters. However, it is worth noting that the Hyades, another famous reference cluster, was found to be inhomogeneous in chemical composition at the 0.02 dex level \citep{2016MNRAS.457.3934L}. Membership determinations in open and globular clusters have dramatically improved with Gaia DR2 \citep{2018arXiv180508726C, 2018A&A...616A..10G}, making them promising validation targets in the future. 

Asteroseismic fields observed by the space missions CoRoT, Kepler,  and K2  are of great interest because stellar surface gravities 
and ages can be determined with very high precision from seismic data \citep[][see also Sect.~\ref{sect:params}]{2013ARA&A..51..353C, 2017ApJ...835...83S}. This valuable information has led several surveys to observe these fields which offer a very good opportunity for calibration. Examples are the APOKASC sample \citep{2014ApJS..215...19P} observing Kepler targets with APOGEE, the K2 stars in RAVE \citep{2017A&A...600A..66V}, and CoRoT targets in GES \citep{2017A&A...598A...5P}. Spectro-seismic datasets also have the potential for inter-comparisons \citep{2017ASInC..14...37J}, provided that surveys agree on stars to observe in common. Some asteroseismic fields also include a few open clusters, which make them even more interesting for reference purposes \citep[the case of M67]{2016ApJ...832..133S}. The use of asteroseismic fields for calibration, training, or validation of automatic pipelines requires the stellar abundances to be determined in those fields with a high level of accuracy and precision. This effort has already started \citep[see, e.g., ][]{2016A&A...594A..43H, 2017A&A...608A.112N}.

%% file: uncertainties_cats.tex
\begin{table}[t]
\hspace{-4cm}
\caption{Uncertainty tests performed by different catalogues and surveys}
\label{tab:errors}
\begin{tabular}{l|ll|lll|ll|l}
\hline
  \multicolumn{1}{c|}{} &
  \multicolumn{1}{c}{Instrument} &
  \multicolumn{1}{c|}{Lines} &
  \multicolumn{1}{c}{Theory} &
  \multicolumn{1}{c}{Params} &
  \multicolumn{1}{c|}{Methods} &
  \multicolumn{1}{c}{Trends} &
  \multicolumn{1}{c|}{External} &
  \multicolumn{1}{c}{Precision} \\
\hline
 {\bf Catalogues} \\
    GBS & yes & yes & yes & yes & yes & yes & yes & yes\\
  Luck & yes & yes & yes & yes & no & yes & yes & no\\
    Bensby & yes & yes & yes & yes & no & yes & yes & yes\\
  AMBRE & yes & yes & no & yes & no & yes & yes & no\\
    APOKASC & yes & yes & no & yes & yes & yes & no & yes\\
HARPS GTO & no & yes & yes & yes & no & yes & yes & no\\
  SPOCS & yes & no & no & yes & no & yes & yes & no\\
\hline
 {\bf Surveys}\\ 
  RAVE & yes & no & no & yes & no & yes & yes & yes\\
  GES & yes & yes & no & yes & yes & yes & yes & no\\
  APOGEE & yes & no & no & yes & no & yes & yes & yes\\
 GALAH & yes & no & yes & no & no & yes & yes & yes\\
 
\hline\end{tabular}
\begin{tabnote}
Notes: Catalogues and surveys are sorted as they appear in Sects.~\ref{sect:catalogues} and \ref{sect:surveys}, except for GBS, which is discussed in Sect.~\ref{sect:gbs}. Columns indicate the uncertainty tests described in Sect.~\ref{sect:errors}. Briefly, Instrument: uncertainty due to different instrumental responses evaluated; Lines: line-by-line abundance dispersion discussed; Theory: 1D-LTE effects assessed; Params: propagation of stellar parameter uncertainties in final abundances; Methods: different methodologies compared; Trends:  consistency of abundances as a function of stellar parameters assessed; External: comparison of results with external sources; Precision: improvement of precision with differential or data-driven methods. 
Reference where information about uncertainties of abundances is found for each catalogue: 
GBS: \cite{GBS4};
Luck: \cite{2018AJ....155..111L}; 
Bensby: \cite{2014A&A...562A..71B};
AMBRE: \cite{2017A&A...600A..22M};
APOKASC: \cite{2016A&A...594A..43H}; 
HARSP GTO: \cite{2012A&A...545A..32A}; 
SPOCS: \cite{2005ApJS..159..141V},\cite{ 2018ApJS..237...38B};
RAVE: \cite{2011AJ....142..193B}, \cite{2017ApJ...840...59C}; GES: \cite{2014A&A...570A.122S}; APOGEE: \cite{2018arXiv180709773H, 2018arXiv180709784J};  GALAH: \cite{2018MNRAS.478.4513B}. 
\end{tabnote}
\end{table}

%% file: surveys.tex
\subsection{Catalogues of stellar abundances from high resolution studies}
\label{sect:catalogues}

\begin{figure}
\includegraphics[scale=0.45]{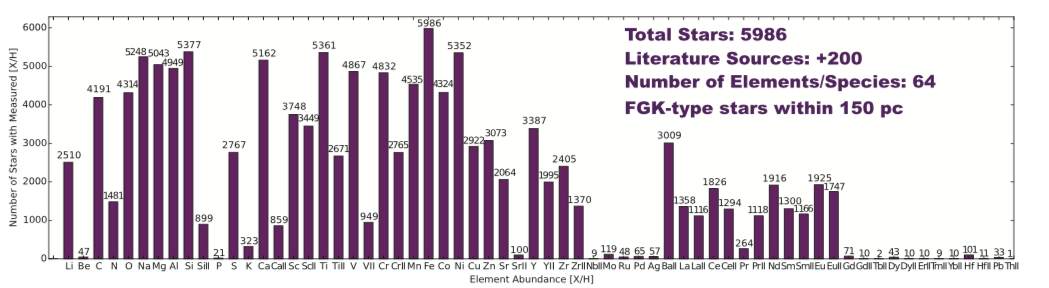}
\caption{Overview of abundances of FGK-type stars in the solar vicinity as included in the Hypatia catalogue. Credit: Hinkel, ApJ, 848, 34, 2017. Reproduced with permission of first author and \copyright\ AAS. } 
\label{4.hypatia}
\end{figure}

\subsubsection{Bibliographic compilations}
\label{sect:compilations}

\cite{2005A&A...438..139S} made an early attempt to combine abundances from different studies, in order to build a large catalogue for the investigation of abundances and kinematic trends in the Galactic disk. This work resulted in 743 stars with abundances of Fe, O, Mg, Ca, Ti, Si, Na, Ni, and Al in the metallicity range $-1.3 < \mathrm{[Fe/H]} < +0.5$, with a typical precision of 0.6 dex. SAGA \citep[Stellar Abundances for Galactic Archaeology Database,][]{2008PASJ...60.1159S} is another compilation of stellar parameters and abundances for $\sim$30 elements from the literature, with the initial motivation to characterise extremely metal-poor stars, in order to constrain the nature of the first stars. The catalogue is now being extended to a larger range of metallicities. It includes more than 1000 stars of the Milky Way and in other nearby galaxies. Large efforts of following-up LAMOST targets are being done in order to homogenise and complete the SAGA database.

The Hypatia catalogue \citep{2014AJ....148...54H, 2017ApJ...848...34H} is a recent compilation which at the time of writing has collected 278\,968 abundance measurements in 171 catalogues for 6156 FGK stars within 150~pc from the Sun. The main purpose is to evaluate the spread in chemical abundances for nearby stars analysed by different groups, which allows one to estimate uncertainties when studying the chemical composition of exoplanet hosts, the connection between thick- and thin-disk stars, or stars with different kinematic properties. The content in terms of stellar abundances is shown in the histogram of Figure~\ref{4.hypatia}. It is seen that light and $\alpha$-elements (C, O, Na, Mg, Al, Si, Ca, Ti), as well as iron-peak elements (Sc, V, Cr, Mn, Fe, Co, Ni) are very common in the literature. Neutron-capture (Sr, Y, Zr, Ba, La, Eu) elements are less common, but still quite popular, as they help to answer important scientific questions regarding stellar and chemical evolution. Other elements have very few abundance measurement in FGK-type stars, for reasons that are discussed in the supplementary text (Sect.~\ref{sup2}). 

\subsubsection{Independent catalogues}
\label{sect:independentcatalogues}

\citet[][and references therein]{2018AJ....155..111L} has undertaken a large high-resolution spectroscopic abundance study. His dataset includes abundances of $\sim$3000 dwarfs, subgiants and giants within $\sim$100 pc from the Sun using good quality spectra selected in public archives of echelle spectrographs. Abundances of C, N, O, Li, Na, Mg, Al, Si, S, Ca, Sc, Ti, V, Cr, Mn, Fe, Co, Ni, Cu, Zn, Sr, Y, Zr, Ba, La, Ce, Nd, Sm, and Eu were determined with a high level of precision. A smaller ($\sim$700 stars), yet very widely used catalogue, was published by \cite{2014A&A...562A..71B}. 
They performed the largest ever ``by-hand'' EW analysis to provide abundances of O, Na, Mg, Al, Si, Ca, Ti, Cr, Fe, Ni, Zn, Y, and Ba for nearby dwarf stars. \cite{2015A&A...577A...9B, 2016A&A...586A..49B} 
complemented the catalogue with Sc, V, Mn, Co, $s-$, and $r-$ process abundances for a subset of the sample. The study has become a reference for how the trends of \xfe\ vs \feh\ are expected to look like for thin and thick disk stars in the solar neighbourhood. 

The AMBRE project consists in the automatic parametrisation of large sets of ESO high-resolution archived spectra from FEROS \citep{2012A&A...542A..48W}, HARPS \citep{2014A&A...570A..68D}, and UVES \citep{2016A&A...591A..81W}. \cite{2016A&A...595A..18G} determined abundances of Li for for 7300 AMBRE stars and \cite{2017A&A...600A..22M} derived Mn, Fe, Ni, Cu, Zn, and Mg abundances for 4666 stars. 

\cite{2016A&A...594A..43H} 
published abundances of C, N, O, Mg, Ca, Si, Ti, S, Al, Na, Ni, Mn, Fe, K, V, P, Cu, Rb, Yb, Co, and Cr for a sample of $\sim$2000 Kepler giant stars which have infrared spectra from APOKASC \citep[see also Sect.~\ref{sect:APOGEE}]{2014ApJS..215...19P}. The stars, as being targeted by Kepler, benefit from asteroseismic data which allow one to better constrain the surface gravity. These data are used to provide a catalogue that is self-consistent, precise, and accurate. 

Planet search programs with radial velocity follow-up are actively generating large spectroscopic catalogues with more than a thousand stars, with high-quality spectroscopy and homogeneous analyses. The characterization of exoplanets requires the properties of the host star to be well-known and this is why several studies have provided the stellar parameters and abundances of the targets of these observing programs. It is worth noting that these programs are dominated by dwarfs. \cite{2012A&A...545A..32A} 
provided chemical abundances of Na, Mg, Al, Si, Ca, Ti, Cr, Ni, Co, Sc, Mn, and V for 1111 FGK-type stars of the HARPS GTO planet search program. While its aim was to characterise planet host stars, this sample has provided insights in stellar populations making it an additional reference for \xfe\ trends as a function of \feh\ in Galactic studies. 
The Spectroscopic Properties of Cool Stars (SPOCS) catalogue \citep{2005ApJS..159..141V} 
contains abundances of Na, Si, Ti, Fe, and Ni for 1040 nearby F, G, and K stars that have been observed by the Keck, Lick, and AAT planet search programs. The California-Kepler Survey \citep[][CKS]{2017AJ....154..107P} is a follow-up program developed to characterise stars with transiting planets detected by Kepler. The catalogue provides HIRES spectra which were analysed by \cite{2016ApJS..225...32B} 
to generate a catalogue of C, N, O, Na, Mg, Al, Si, Ca, Ti, V, Cr, Mn, Fe, Ni, and Y abundances. 
The catalogue of Spectroscopic Parameters and atmosphEric ChemIstriEs of Stars \citep[][SPECIES]{2018A&A...615A..76S} 
is built from public spectra and includes Na, Mg, Al, Si, Ca, Ti, Cr, Mn, Ni, Cu, and Zn abundances for about 1000 planet host stars. Note that the catalogues above have a significant number of stars in common, allowing for comparisons. 

\subsection{Chemical abundances of spectroscopic surveys}
\label{sect:surveys}
The catalogues presented above have shown that chemical abundances of FGK-type stars provide key information on their formation process and site. The next step is to construct large chemical maps of the Galaxy to constrain models of its formation and evolution. In addition, stars need to have a well-defined selection function in order to probe Galaxy models properly. To that aim, stars are surveyed with multi-object spectrographs over several years, and automatic pipelines to measure abundances are designed. 

The first efforts in the field of massive spectroscopy were dedicated to the search of metal-poor stars. This started with large objective-prism surveys that produced hundreds of candidates followed-up at medium or high resolution. This pioneering work, essential to the later expansion into the era of industrial abundances, is reviewed by \cite{1985AJ.....90.2089B}. 
The first very ambitious project aiming at determining spectroscopic abundances at industrial scales is the Sloan Digital Sky Survey (SDSS) with its Sloan Extensions for Galactic Understanding and Exploration \citep[][SEGUE]{2009AJ....137.4377Y}, 
which provided moderate-resolution ($\sim$1800) spectra for well over 500\,000 unique stars. The SEGUE Stellar Parameter Pipeline \citep[SSPP, ][]{2008AJ....136.2022L} 
was developed specifically to obtain large-scale estimates of \teff, \logg, and [Fe/H], and was later extended to [C/Fe] and [$\alpha$/Fe] determinations. The SSPP pioneered the use of multiple techniques to determine stellar parameters, as well as the validation with open and globular clusters \citep{2008AJ....136.2050L}
and the estimation of errors by comparison to parameters from high-resolution studies \citep{2008AJ....136.2070A}.
The impact of SDSS on the understanding of the Milky Way stellar populations is reviewed in \cite{2012ARA&A..50..251I}. 

\begin{figure}
\includegraphics[scale=1.0]{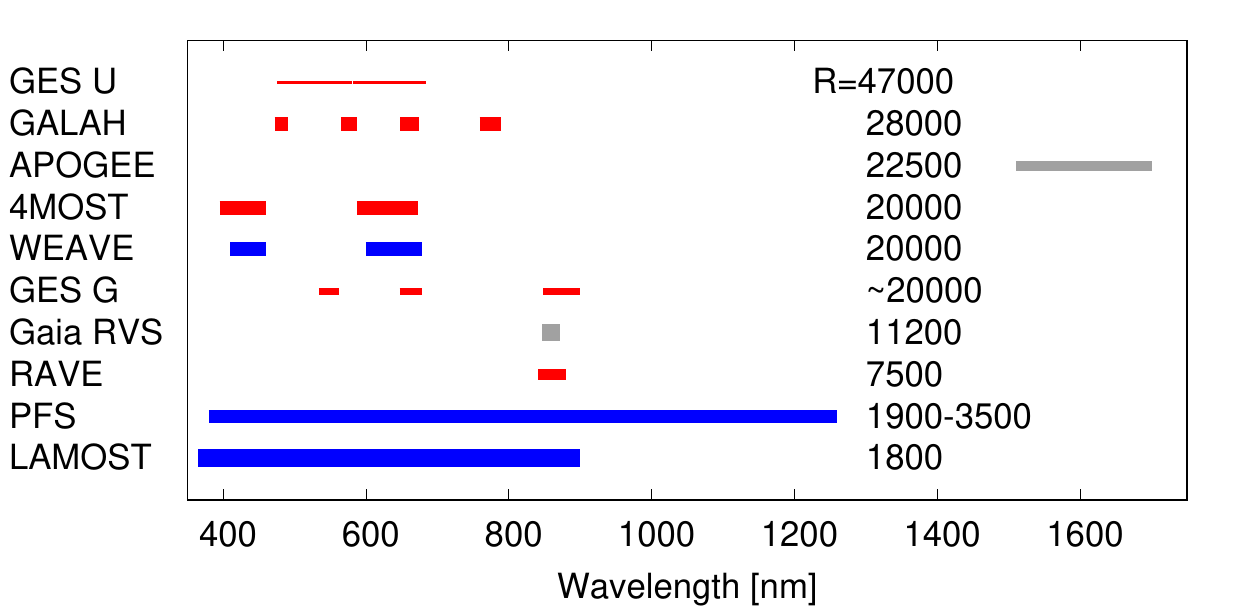}
\caption{Overview of on-going and future spectroscopic surveys (see Sect.~\ref{sect:surveys}, GES U/G = Gaia-ESO UVES/GIRAFFE, respectively), sorted by spectral resolution (see labels to the right). Horizontal lines show the covered wavelength intervals. Northern, southern, and all-sky surveys are represented by blue, red, and grey colours, respectively. Broader lines indicate larger numbers of target stars.} 
\label{fig:wave_surveys}
\end{figure}

In this section, we describe the main on-going and future spectroscopic surveys which deliver abundances of at least 5 individual elements. We discuss the targeted accuracy, methods, performances, and calibration strategies that they implement in their latest data releases following our listing in Table~\ref{tab:errors} for uncertainties of the abundances. Their main characteristics are shown in Figure~\ref{fig:wave_surveys}.
The abundances determined by each of the on-going surveys are marked in the periodic table of Figure~\ref{fig:periodic_table_surveys} for each survey with a different colour.

\subsubsection{RAVE}
\label{sect:RAVE}
The RAdial Velocity Experiment \citep[RAVE,][]{2006AJ....132.1645S} was the first large survey that provided abundances of several individual elements. The observations were performed with the 6-degree field multi-object spectrograph on the 1.2~m UK Schmidt Telescope of the Anglo-Australian Observatory. DR5 \citep{2017AJ....153...75K} contains 457\,589 stars in the magnitude range of $9<I<12$, observed between 2003 and 2013. The spectra are not public. 

The chemical pipeline is described in \cite{2011AJ....142..193B} 
and obtains abundances of Mg, Al, Si, Ca, Ti, Fe, and Ni for about 300\,000 stars based on EWs (cf. Sect.~\ref{sect:analysis_methods})  from 604 absorption lines identified in spectra of the Sun and Arcturus.
Stellar parameters are derived from the spectra by fitting to a grid of synthetic spectra, which was built using MARCS models, the Turbospectrum radiative transfer code, and astrophysical $gf$-values. 
The best model is found from a combination of a decision-tree algorithm
\citep[DEGAS, ][]{Bijaoui_2012} and a projection method of models to data \citep[MATISSE, ][]{2006MNRAS.370..141R}. 
The final parameters are calibrated by a modification of the parameters with second-order polynomials obtained from the comparisons against samples of reference stars. Metallicities are calibrated considering sets of reference stars which are a combination of literature sources of results obtained with high-resolution optical spectra \citep[PASTEL, \gbs s,][among others]{2005A&A...438..139S}. Surface gravities are further calibrated using 72 giants with seismic information from K2 \citep{2017A&A...600A..66V} and \teff\ is calibrated with photometric relations. To assess the errors in stellar parameters, extensive comparisons with external datasets are performed (see below). 

{\bf Abundance uncertainties:}
{\it Random/Instrument:} synthetic spectra with different levels of added noise and repeated observations.
{\it Random/Lines:} Not reported.
{\it Systematic/Theory:} Not reported.
{\it Systematic/Params:} effect on abundances from uncertainties in parameters.
{\it Systematic/Methods:} Not reported.
{\it Bias/Trends:} investigated as a function of temperature. 
{\it Bias/External:} comparison with \citet[]{2005A&A...438..139S, 2011ApJ...737....9R}.
{\it Final error budget:}
The combination of these different tests showed the typical accuracy to be 0.2~dex depending on S/N and atmospheric parameters. 

Together with the standard DR5 pipeline, RAVE spectra have been re-analysed by \cite{2017ApJ...840...59C} with the data-driven method {\it The Cannon} \citep{2015ApJ...808...16N}, providing abundances of O, Mg, Al, Si, Ca, Fe, and Ni for red giant stars in a complementary catalogue called RAVE-on. The training set was built with RAVE stars that are also in APOGEE (see Sect.~\ref{sect:APOGEE}).
The typical precision, estimated from repeated observations is much better (0.07~dex) than for the standard pipeline, but the sample is smaller due to the challenge of finding suitable training sets for the entire parameter space covered by RAVE.

\cite{2017AJ....153...75K} 
compare stellar parameters of stars in common between RAVE and other surveys. Their Table 5 is a nice summary of the situation, showing, in general, agreements for S/N $>$ 50 spectra in \teff\ on the order of 100~K, and in \logg\ and \feh\ of about 0.1 to 0.2~dex. Recent comparisons of the different sets of RAVE parameters (DR4, DR5, RAVE-on) to independent determinations by \cite{2018AJ....155..256P} 
in the low metallicity regime  show significant discrepancies well above the errors mentioned above. \cite{2017ApJ...840...59C} compare the RAVE-on abundances with GES (see below) for 30 stars in common. The total differences vary between 0.06~dex for [Al/H] and [Mg/H] to 0.26~dex for [Si/H]. 

\subsubsection{Gaia-ESO Survey}
\label{sect:GES}


The Gaia-ESO public spectroscopic survey \citep[GES,][]{2012Msngr.147...25G,2013Msngr.154...47R} targets $10^5$ stars in different populations of the Milky Way, as well as in a large sample of open clusters of different characteristics with FLAMES on the 8~m VLT. One special feature of this survey is that it targets stars of wider spectral ranges than FGK-type owing to different science cases for which groups within the consortium have developed their own analysis methods. The GIRAFFE spectrograph was used with several wavelength ranges, depending on the stellar type, providing spectra for stars down to $V$= 19. In parallel, UVES spectra were obtained in each field for brighter stars. Observations took place between 2011 and 2018. The latest data release (DR3) includes observations and data processing between 2011 and 2014. The information about the release can be found in an ESO document\footnote{\url{http://www.eso.org/rm/api/v1/public/releaseDescriptions/92}}. In total 12 FLAMES configurations were used, the wavelength intervals and resolutions of which are given in \cite{2017A&A...598A...5P}, and Fig.~\ref{fig:wave_surveys} shows those used for FGK-type stars. 
Different working groups share the spectral analysis task depending on the spectral type. 
Within each of them several teams participate in the analyses. For FGK-type stars, they span the entire range of methods described in Sect.~\ref{sect:analysis_methods}, namely from fitting synthetic spectra to determination of EWs. In the latest data release, abundances of Li, C, N, O, Na, Mg, Al, Si, Ca, Sc, Ti, V, Cr, Fe, Co, Ni, Zn, Y, Zr, Ba, La, Ce, Nd, and Eu are included. 

Common inputs were adopted for the analyses, including a set of MARCS model atmospheres 
and a grid of synthetic spectra computed with Turbospectrum following  \cite{2012A&A...544A.126D}. 
Several teams did not employ the grid of model spectra, and used other radiative transfer codes such as SME or MOOG. For the line list, the best atomic data available in the literature were collected, excluding astrophysically calibrated $gf$-values \citep{heiter-linelist}. In addition, lines used to determine abundances were flagged according to their laboratory data quality, as well as according to the amount of blending estimated from a comparison of observed and synthetic line profiles for the Sun and Arcturus. Teams were encouraged to use this flagging system to choose the best lines for their analysis. However, this did not impede a different line selection among them (see Fig.~\ref{fig:heatmap} in Sect.~\ref{sect:line_selection}).

{\bf Abundance uncertainties:}
{\it Random/Instrument:} Repeated observations of the \gbs s, other observations across S/N, and cluster stars.
{\it Random/Lines:} For the UVES analysis, the line-by-line scatter is considered to perform line selection for final results. Furthermore, if a method does not treat HFS properly for a line that requires it, their result for that line is rejected.
{\it Systematic/Theory:} The SME method included non-LTE calculations for some elements for UVES spectra. They are, however, weighted in the same way as other methods.
{\it Systematic/Params:} Cluster stars to assess scatter. Final results on abundances are weighted according to their errors in stellar parameters. 
{\it Systematic/Methods:} Most of the decisions which determine the final abundances are based on agreement between methods.
{\it Bias/Trends:} Trends of the abundances derived from each line by all methods are checked as a function of \teff\ and \logg. Each line is excluded from or included in the final result depending on the trend and scatter. Final abundances are checked with cluster stars.
{\it Bias/External:} The abundances of the benchmark stars and of the Sun \citep{2007SSRv..130..105G} are used as priors to help establishing the scales. No other catalogue is used to compare abundances.
{\it Final error budget:} Estimated from a Bayesian modelling to infer the typical errors of the parameters and abundances from the distributions of values provided by the teams, and is typically between 0.1 and 0.2 dex, depending on S/N and spectral type.
We note that published abundances are provided for stars selected to have reliable parameters only, namely with uncertainties of less than 5\% in \teff, 0.4~dex in \logg, and 0.2~dex in \feh. These criteria resulted in 2000 to 9000 stars with abundances in DR3, depending on the element.

Due to the many different methodologies, spectra, and stellar types targeted by the survey, GES dedicates substantial efforts to understanding any systematic differences. A complex communication strategy has been put in place to provide feedback between the data reduction and the analysis teams, as well as the homogenization group producing the final abundances. The size of the data set increases with each data release, and the abundances improve thanks to the various inter-comparisons of results between teams and spectral setups the survey considers.

The final data release will include spectra of all observations, the stellar parameters will use the {\it Gaia}-DR2 astrometric information as priors, and the calibration strategy will include the \gbs s as well as a sample of seismic targets from CoRoT and K2. Everything (individual results from each methodology and spectra) will be made publicly available. 

\subsubsection{APOGEE}
\label{sect:APOGEE}

The Apache Point Observatory Galactic Evolution Experiment,  \citep[APOGEE,][]{2017AJ....154...94M}, one of the SDSS surveys, was optimized to explore
the “dust-hidden” populations in the Milky Way. Using the 2.5~m Sloan Telescope, APOGEE has been collecting spectra since 2011. Recently, an identical spectrograph has been installed at the 2.5~m du Pont Telescope in Chile with the goal to extend the survey to the Southern Hemisphere. Observations on that telescope started in February 2017. Spectra cover the range from 1.514 to 1.696 microns at R$\simeq$22\,500. The latest SDSS data release (DR14) described by \cite{2018ApJS..235...42A} corresponds to $\sim$260\,000 stars observed until 2016. The release includes abundances of C, N, O, Na, Mg, Al, Si, P, Si, K, Ca, Ti, V, Cr, Mn, Fe, Co, Ni, and Rb. The stars are predominantly red giants and other luminous post-main-sequence stars situated in the obscured parts of the Galactic disk and bulge. 

The APOGEE Stellar Parameter and Chemical Abundances Pipeline, ASPCAP, is described in \cite{2016AJ....151..144G}. ASPCAP is based on the FERRE code (see Table~\ref{tab:tools}), which finds the best fit between oberved and synthetic spectra.
\cite{2018arXiv180709773H} 
describe the details of the spectral analyses of the latest data releases of APOGEE (DR13 and D14).
The synthetic grid was built with ATLAS9 models (MARCS models are included for the coolest M giants) and Turbospectrum and has many dimensions (\teff, \logg, [M/H], \afe, \vmic, \vmac; [C/H], [N/H] for giants; rotational velocity for dwarfs).
Micro- and macroturbulence are determined from empirical relations that depend on stellar parameters. The final stellar parameters are then empirically calibrated.
Similar to RAVE, \teff\ is calibrated with the help of photometric temperatures and \logg\ with stars which benefit from seismic observations. The latter uses $\sim$2000 stars from the APOKASC catalogue \citep{2014ApJS..215...19P}, 
a joint effort between APOGEE and Kepler for the purpose of this calibration.

Elemental abundances are determined by fitting parts of the spectra within spectral windows located around features of each element. These windows are constructed using weights for each spectrum pixel proportional to the change of the flux with the abundance at the corresponding wavelength.
The line list employed comes mainly from NIST, and is described in \cite{2015ApJS..221...24S}. 
For the latest data releases it has been improved by adding HFS for Al and Co, as well as molecular data for H$_2$O, which is important for the coolest stars. As laboratory data for IR lines are more scarce than in the optical, $gf$-values and damping constants were astrophysically calibrated (see Sect.~\ref{sect:line_data}) for $\sim$20\,000 lines by fitting synthetic line profiles to observed ones for the Sun and Arcturus.  

{\bf Abundance Uncertainties:}
{\it Random/Instrument:} Evaluated using stars in 23 globular and open clusters; scatter in abundances is provided as a function of S/N.
{\it Random/Lines:} Since abundances are determined from a simultaneous fit of all absorption features, this uncertainty is not given. However, differences compared to optical regions are studied in \cite{2018arXiv180709784J}.
{\it Systematic/Theory:} Not reported.
{\it Systematic/Params:} Evaluated using clusters; scatter in abundances is provided as a function of stellar parameters.
{\it Systematic/Methods:} Only one method used.
{\it Bias/Trends:} Investigated using clusters, finding trends as a function of metallicity and temperature. Abundances are calibrated with polynomials as a function of metallicity and temperature.
{\it Bias/External:} [X/M] was calibrated shifting the zero point to force the mean abundance ratios of all stars with $-0.1<\mathrm{[M/H]}<0.1$, $-5^\circ < b < 5^\circ$ and $70^\circ < l < 110^\circ$ to have solar abundance ratios based on the catalogues of \cite{2014A&A...562A..71B}. 
{\it Final error budget:} 
For stars with \teff=4500~K, [M/H]=0, and S/N=100 typical uncertainties vary between 0.02 and 0.1~dex. The global uncertainty (the scatter of all cluster stars) is about 0.02~dex larger. 

APOGEE is the first ambitious project to collect near-IR spectra at massive scales, opening a new window of spectroscopy and pushing the progress in modelling of spectral lines which for a long time have been essentially unexplored. The spectra have, in general, S/N$>$100, and are public, which has produced several complementary datasets of parameters and stellar abundances with alternative methods. APOGEE data have became a favourite playground for developing new machine-learning tools to derive abundances at industrial scales. 
The first application has been {\it The Cannon} \citep{2015ApJ...808...16N}. The results on abundances of a modified version \citep{2016arXiv160303040C} are part of DR14, which uses a subset of ASPCAP labels to train the model for providing {\it The Cannon} labels. Comparisons of the results from {\it The Cannon} and ASPCAP can be found in \cite{2018arXiv180709773H}. 

{\it The Payne} has been introduced by \cite{2018arXiv180401530T}, 
deriving parameters and 15 elemental abundances from APOGEE spectra. 
The method fits the data to models in a special way which allows precise and quick determination of many labels simultaneously. It is based on neural networks with gradient spectra (change in model spectra as each stellar label is varied by a small amount). Its performance in terms of abundance precision is competitive with {\it The Cannon}, but offers the possibility to build a parameter-complete training set since it is based on synthetic spectra. 
\cite{2018arXiv180804428L} 
showed how abundances of APOGEE spectra can be derived with deep learning
using artificial neural networks (ANNs). The results, trained on a selection of reliable ASPCAP labels, are as precise as those from {\it The Cannon}. Stars are also analysed extremely quickly showing the potential of this method for future big datasets of stellar abundances. 

An early attempt to compare parameters and $\alpha$-element abundances of $\sim$200 stars in common between GES and APOGEE can be found in \cite{2017ASInC..14...37J}, 
who discuss the main outputs of a workshop held with key developers of survey chemical pipelines. \teff, \logg, \feh, and \afe\ agree within 120~K, 0.27, 0.15, and 0.14~dex, respectively. However, they explain that \afe\ should not be directly compared (or merged between surveys!), since the values from GES and APOGEE are based on absorption features that are produced by different $\alpha$-elements.

\cite{2018arXiv180709784J}, 
compared stellar parameters and abundances from APOGEE 13 and 14, determined with the ASPCAP pipeline and by {\it The Cannon}, with independent analyses focused on the optical. They selected five studies in the literature with high-quality parameters and abundances that had at least 100 stars in common with APOGEE, including GES. For most of the elements (C, Na, Mg, Al, Si, S, Ca, Cr, Mn, Ni), the DR14 ASPCAP analysis showed systematic differences to the comparison samples of less than 0.05~dex (median), and random differences of less than 0.15~dex (standard deviation). Fe, Mg, and Ni are the elements which show the best agreement with the reference values.

\subsubsection{GALAH}
\label{sect:galah}
The Galactic Archaeology with HERMES (GALAH) survey is a large high-resolution spectroscopic
survey using the High Efficiency and Resolution Multi-Element
Spectrograph (HERMES) on the 3.9~m Anglo-Australian Telescope. The HERMES spectrograph
provides spectra for $\sim$400 stars simultaneously over a 2 degree field of view \citep{2015MNRAS.449.2604D}. The goal is to observe up to $10^6$ stars and to measure 30 individual chemical element abundances per star from Li to Eu with errors below 0.1~dex.

GALAH DR2 \citep{2018MNRAS.478.4513B} provided abundances of Li, C, O, Mg, Si, Ca, Ti, Na, Al, K, Sc, V, Cr, Mn, Fe, Co, Ni, Cu, Zn, Rb, Sr, Y, Zr, Ba, La, Ru, Ce, Nd, and Eu 
for 342\,682 stars observed between January 2014 and 2018. The spectra are not public. The pipeline has two steps: (1) A training set is defined which is analysed with SME and (2) {\it The Cannon} is applied to the entire dataset. 
The spectral analysis of the training set considers MARCS models and the Gaia-ESO line list (see Sect.~\ref{sect:GES}), complemented by following the same procedure for the spectral ranges outside GES. An interesting characteristic of this survey is that the training set has been analysed in non-LTE (using SME), for elements for which this is possible (Li, O, Na, Mg, Al, Si, and Fe). The training set was built in order to be representative of the parameter space, with stars having relevant information from the literature, such as the \gbs s, asteroseismic targets, stars with known parallaxes, members of open and globular clusters, and further stars used as reference in other projects (TESS, K2, APOGEE). This resulted in a total of 10\,605 stars, although not all of them have abundance measurements for all elements. Validation tests include repeated observations, \gbs s, open and globular clusters, and K2 asteroseismic targets. To assess any biases, GALAH performs leave-out tests, i.e., randomly selecting 80\% of the stars from the training set for training, and testing on the other 20\%. For the production run however the entire training set is used.


{\bf Abundance uncertainties:}
{\it Random/Instrument:} Analysis of repeated observations in the field and scatter of abundances found in M67. 
{\it Random/Lines:} Since SME performs a simultaneous fit of all lines, this is not reported. 
{\it Systematic/Theory:} The SME method performs non-LTE calculations for some elements. Comparisons are reported in \cite{2018arXiv180405869B} and \cite{2018MNRAS.tmp.2299G}.
{\it Systematic/Params:} Not reported.
{\it Systematic/Methods:} Not reported.
{\it Bias/Trends:} Investigated as a function of \logg.
{\it Bias/External:} Investigated using leave-out tests.
{\it Final error budget:} Estimated with stars belonging to the open cluster M67. Uncertainties were found to range from the
highest precisions of 0.04--0.08 dex (Fe, Al, Sc, Ti, V, and Cu), over high precisions of 0.08--0.12 dex (C, Na, Si, Cr, and Mn), and intermediate precisions of 0.12--0.16 dex (O, Mg, K, Ca, Co, Ni, Zn, and Y) to low precisions $>$0.16 dex (Li, Ba, La, and Eu).

\subsubsection{LAMOST}

The LAMOST Experiment for Galactic Understanding and Exploration (LEGUE) survey of  Milky Way stellar structures is conducted at the 4~m Guo Shoujing Telescope in China \citep{2006ChJAA...6..265Z, 2012RAA....12..735D}. Using a modified Schmidt telescope, LAMOST can observe up to 4000 objects simultaneously over a 20 square degrees field-of-view. LAMOST DR3 published spectra for more than 3 million stars. Despite a low resolution (see Fig.~\ref{fig:wave_surveys}), several groups have managed to measure elemental abundances from this vast set of spectra. \cite{2016RAA....16..110L} developed a template-matching technique to measure [$\alpha$/Fe] ratios with an accuracy better than 0.1 dex for S/N $>$ 20. With {\it The Cannon}, \cite{2017ApJ...836....5H} performed a label transfer from APOGEE to LAMOST and measured [$\alpha$/M] for 454\,180 giants, 20\% of the LAMOST DR2 and the largest catalogue of [$\alpha$/M] for giant stars to date. \cite{2017MNRAS.464.3657X} estimated stellar atmospheric parameters, absolute magnitudes and elemental abundances ([M/H], [$\alpha/M$], [C/H], [N/H]) from the LAMOST spectra with Kernel-based principal component analysis, using an algorithm trained with stars in common with other catalogues (Hipparcos, Kepler, APOGEE). They obtained a precision of 0.1 dex for [Fe/H], [C/H], and [N/H], and better than 0.05 dex for [$\alpha$/M]. \cite{2018AJ....155..181B} obtained [$\alpha$/H] abundances for 1\,097\,231 stars. The typical precision is $\sim$0.15 dex in [Fe/H], and $\sim$0.1 dex in [$\alpha$/Fe] for spectra with S/N $>$ 40, with some differences between dwarf and giant stars. \cite{2017ApJ...849L...9T} measured 14 elemental abundances (C, N, O, Mg, Al, Si, Ca, Ti, V, Cr, Mn, Fe, Co, Ni) for objects with S/N $>$ 30 
using {\it The Payne}, with a training set made of $\sim$500 cross-matched objects between the APOGEE DR13 and LAMOST DR3 catalogues with a LAMOST S/N $>$ 200. 
The typical precision is 0.1~dex. 

\cite{2015AJ....150..187L} 
applied a special version of the SSPP to LAMOST spectra (LSSPP), thus obtaining [$\alpha$/Fe] and [C/Fe], and compared their results to the parameters obtained with the regular LAMOST pipeline, and to those from RAVE, APOGEE, and SEGUE. SEGUE and LAMOST are found not to be on the same abundance scale, with an offset of 0.15 in metallicity. APOGEE and SEGUE are in very good agreement for \afe, while LSSPP seems to underestimate it.

The parameters of APOGEE and LAMOST have also been compared by \cite{2018arXiv180707625A}. Using $\sim$40\,000 stars in common between APOGEE DR14 and LAMOST DR3, they evaluated mean [Fe/H] discrepancies as a function of S/N, \teff, [Fe/H], and \logg. Even though on average the metallicity offset between the two surveys was found to be low (0.03 or 0.06~dex depending on the LAMOST pipeline used) and the scatter reasonable (0.13~dex), complex dependencies between the parameters were found. They report significant discrepancies of 0.10 to 0.15~dex among metal-poor stars, and also show that the differences increase with decreasing \teff.

LAMOST is the perfect dataset to test the new generation of data-driven methods, because, on the one hand, the spectra are of low resolution and do not allow one to perform the standard methods for abundance determination. On the other hand, the dataset is very large, so fast methods are needed. 

\begin{issues}[INTERCOMPARISONS OF SURVEYS]
Due to their different characteristics, the spectroscopic surveys and catalogues have only several hundreds of stars in common, or a few thousands for the largest ones. Their intercomparison is mandatory to track systematic differences, understand their origin and put all survey products onto the same scale. 

One primary ambition in the field is to have stellar properties and abundances on the same scale, as this allows the community to straightforwardly merge the datasets from different surveys for their specific science case. This is particularly crucial when the samples are chosen from the Gaia database, which covers the entire sky. The efforts made by the surveys to calibrate their stellar parameters using benchmark objects are devoted to that goal. However, the situation is not yet satisfactory, because systematic differences become apparent when comparing the different surveys and catalogues. For the time being, this is preventing the community to make optimal use of the huge chemical information that is available. A few systematic studies comparing surveys have been published recently \citep[e.g.,][]{2017ASInC..14...37J, 2018arXiv180707625A, 2018arXiv180709784J}, and we expect that more such studies will become available in the coming years.
\end{issues}

\subsubsection{Forthcoming industrial abundances}
The era of large spectroscopic surveys has just begun. Several even bigger projects are planned for the next decade. The next future survey is WEAVE \citep{2012SPIE.8446E..0PD}, a new multi-object survey spectrograph for the 4.2~m William Herschel Telescope (WHT) at the Observatorio del Roque de los Muchachos, on La Palma (Canary Islands). The facility will be capable of obtaining about 1000 spectra over a two-degree field of view in a single exposure starting in 2019. WEAVE's fibre-fed spectrograph comprises two arms, one optimised for the blue and one for the red, and offers two possible spectroscopic resolutions, 5000 and 20\,000. 

Gaia has a spectrograph on board (RVS) covering a wavelength interval around the Ca IR triplet with a resolving power of $R\sim$11\,500, similar to the RAVE or the GIRAFFE HR21-setup spectra \citep{2016A&A...585A..93R}. We know from RAVE that, from such spectra, it is possible to derive abundances for a limited number of elements (see Sect.~\ref{sect:RAVE}). The third Gaia data release, expected in the first half of 2021,
will release millions of stellar parameters, abundances and spectra\footnote{\url{ https://www.cosmos.esa.int/web/gaia/release}}.

The 4-metre Multi-Object Spectroscopic Telescope project \citep[4MOST,][]{2016SPIE.9908E..1OD} is the next 
ESO spectroscopic survey facility on the VISTA telescope,  scheduled to start observations in 2022. With its large field-of-view it will be able to simultaneously obtain spectra of $\sim$2400 objects. \cite{2018IAUS..334..225F} present an overview of the science goals, spectral properties, and the design of the chemical pipeline.
From the high-resolution spectra it will be possible to measure chemical-abundance ratios to better than 0.1~dex for Fe, Mg, Si, Ca, Ti, Na, Al, V, Cr, Mn, Co, Ni, Y, Ba, Nd, and Eu, and better than 0.2~dex for Zr, La, and Sr \citep{2013AN....334..197C}. 
This precision comes from the number of lines in simulated spectra at different S/N, and exclude systematic uncertainties related to stellar parameters or atomic data. Their Table~1 shows the number of lines for each element that is expected to be detected with 4MOST. 
However, \cite{2015AN....336..665H} 
discuss the possibility that, from the bluest arm of 4MOST, it might be possible to detect new elements for certain stars, these being heavy elements such as Pb, Th, Dy, Ce, and Sm. Similar abundances should be obtained from WEAVE due to their spectral similarities (see Fig.~\ref{fig:wave_surveys}).
4MOST will be succeeded by another ESO facility on the 8~m VLT, the Multi-Object Optical and Near-infrared Spectrograph \citep[MOONS,][]{2014SPIE.9147E..0NC}. It will combine a wide field of view ($\sim$500 square arcmin) with a large degree of multiplexity and wavelength coverage (1000 fibers, optical to near-IR). MOONS has a medium-resolution ($R=$ 5000) and a high-resolution ($R=$ 20\,000) mode, the latter focused on the J and H bands. 

The Prime Focus Spectrograph \citep[PFS]{2014PASJ...66R...1T} is the next generation facility instrument on the 8.2~m Subaru Telescope. It is a very wide-field, massively multiplexed optical and near-infrared spectrograph which will dedicate a portion of its time to observe $10^6$ stars in the Galactic thick disk, halo, and tidal streams for magnitudes down to V $\sim$ 22. A medium-resolution mode with $R=$ 5000 to be implemented in the red arm will enable the measurement of $\alpha$-element abundances.
Finally, the Maunakea Spectroscopic Explorer \citep[MSE]{2016arXiv160600043M}, a rebirth of the 3.6~m Canada-France-Hawaii Telescope on Maunakea, is a proposed 11.25~m wide-field (1.5 square-degree) telescope, equipped with multi-object spectrographs, that will obtain for each pointing more than 4000 optical and near-infrared spectra of low, intermediate, and high resolution. 

 



%% file: sect_common_elements.tex
\begin{figure}
\includegraphics[scale=0.45]{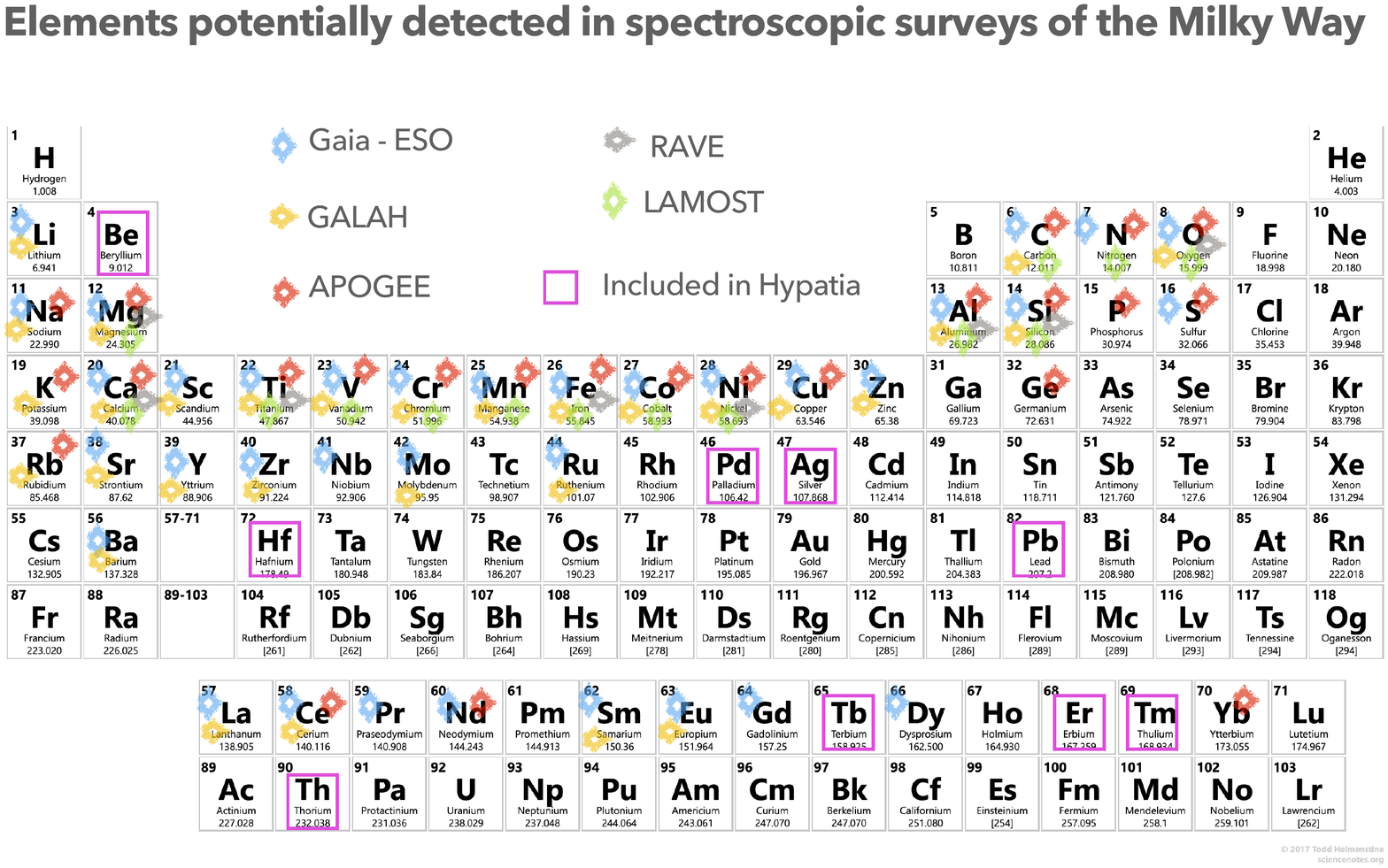}
\caption{Periodic table annotated with the surveys for which lines of a given element can be detected. Elements in squares are included in Hypatia, but are not planned to be studied in spectroscopic surveys.} 
\label{fig:periodic_table_surveys}
\end{figure}

Widely measured elements can be identified in the histogram of Figure~\ref{4.hypatia} as those that have more than $\sim$4000 measurements and are covered by at least all surveys with high resolution and large wavelength coverage (GALAH, GES, and APOGEE, see Figure~\ref{fig:periodic_table_surveys}). 
These are C, O, Na, Mg, Al, Si, Ca, Ti, V, Cr, Mn, Fe, Co, and Ni. Cu and Nd are common elements targeted by surveys but not as common as the others in Hypatia.

To evaluate the precision of abundances for the common elements (see Section \ref{sect:common_elements}), let us take as an example the abundances published by GES to illustrate the situation in the optical wavelength region. We choose GES because of its unique strategy of combining several results using different methods based on a common list of atomic data with special effort in flagging ``good" and ``bad" lines (see Sect.~\ref{sect:GES} and \citealt{heiter-linelist}). 
This allows one to assess the method-to-method or line-to-line dispersion (MMD or LLD, respectively) for different species.
The precision of the abundances evaluated using the MMD was found to be highest ($<0.15$ to $0.2$~dex) for the species
\nai, \mgi, \ali, \sii, \cai, \scii, \tii, \vi, \fei, \zri, \moi, and \baii\
\citep[p.~23]{2014A&A...570A.122S}. 
The elements with precise abundances in common among all setups are \ali, \sii, and \cai.
In the IR, abundances more precise than 0.05~dex in APOGEE DR14 \citep{2018arXiv180709773H} are those of C, O, Mg, Al, Si, Ca, Ti, Cr, Mn, Fe, and Ni.
In summary, popular elements in spectroscopic surveys in the optical and IR that are also precise are Mg, Al, Si, Ca, Ti, and Fe.  These elements all have agreements and biases between optical and IR surveys within the uncertainties \citep{2018arXiv180709784J}.  
The works of \cite{GBS4}, \citet{heiter-linelist}, \cite{2016A&A...594A..43H}, 
and of \cite{2018arXiv180709784J} help us to investigate in depth which are the best lines used for abundance analyses in the IR and optical.  We exclude Fe from this discussion as being the element with the largest number of lines available in FGK-star spectra, which forms the basis for the metallicity parameter.

\begin{marginnote}[120pt]
\entry{Magnesium}{6318\AA, 6319\AA, 15231\AA, 15693\AA, 15740\AA, 15879\AA, 15954\AA, 16365\AA }
\entry{Aluminum}{5557\AA, 6696\AA, 6699\AA, 8773\AA, 8774\AA,  16763\AA}
\entry{Silicon}{5690\AA, 5701\AA, 5949\AA, 15377\AA, 15888\AA, 16216\AA, 16681\AA,  16828\AA}
\entry{Calcium}{5867\AA, 16150\AA, 16157\AA,  16197\AA}
\entry{Titanium}{5689\AA, 5702\AA, 6091\AA, 15873\AA}
\end{marginnote}

\begin{summary}[MOST PRECISE AND ACCURATE ELEMENTS]
\begin{itemize}
\item {\bf Magnesium} In the GES range, twelve \mgi\ lines are included. Among them, two have accurate atomic data, and are good lines for a variety of stars and methods. In APOGEE there are 14 \mgi\ lines, among which six are suitable for a large variety of stars. Non-LTE corrections have been determined in the optical by \cite{2015A&A...579A..53O},  and in the IR by \cite{2017ApJ...835...90Z} and \cite{2017ApJ...847...15B}.
\item {\bf Aluminum} No reliable experimental transition probability data exist for the five \ali\ lines in the GES line list.
However, none of them is considered heavily blended, possibly explaining the good precision (MMD) achieved despite the lack of reliable laboratory data. They can be used for a variety of stars, preferentially of solar metallicities. 
In APOGEE, three \ali\ lines are present, although one  seems to be more robust for different types of stars. Non-LTE studies of optical and IR lines have been performed by \cite{2017A&A...607A..75N}.  
\item {\bf Silicon} Among the 45 \sii\ lines visible in GES spectra, only three  have good atomic data and are blend-free in dwarfs and giants. The partially blended line at $\lambda$5708\AA\ with good atomic data is strong enough to be measured in metal-poor stars. APOGEE has 17 \sii\ features, among which 5 are good lines to be used for a variety of stars. Improved modelling for \sii\ lines has been done by \cite{2012ApJ...755...36S} and \cite{2016ApJ...833..137Z} for optical and IR lines. Non-LTE corrections in the optical are provided for the \gbs s by \cite{GBS4}.
\item {\bf Calcium} The GES has 31 \cai\ and eight \caii\ lines detectable, among which twelve and two have reliable laboratory data, respectively. The lines are highly sensitive to stellar parameters and therefore, only one  can be used to derive abundances for a large variety of spectral types (excluding metal-poor stars). In APOGEE, four lines are visible, but only three can be used for a wide range in stellar parameters.  Non-LTE corrections in the optical are provided for the \gbs s by \cite{GBS4}. In the IR calculations are needed. 
\item {\bf Titanium} After \fei, \tii\ is the element with the most numerous absorption features, including 105 in GES spectra, among which 23 have good atomic data and are largely blend-free. Three lines  can be used for elemental abundances in a large variety of stars. In APOGEE nine Ti lines are detected (but see Sect.~\ref{sect:line_selection} for a discussion on their applicability). Non-LTE effects have been studied by \cite{2011MNRAS.413.2184B} for optical lines.  In the IR calculations are needed. 
\end{itemize}
\end{summary}

Common elemental abundances which can be precisely measured from IR spectra, but not from optical spectra, are C, O, and Mn. C and O abundances can be derived from molecular features, while few clean Mn lines are suitable. In the optical the situation is more complicated,
as there is only one \ci\ and one [\oi] line which have good atomic data and are free of blends. The \ci\ line is very weak and [\oi] lies in telluric regions. The O triplet at 7770\AA\ is popular, and is included in GALAH, but it is subject to strong non-LTE effects \citep{2016MNRAS.455.3735A}. For Mn, although many lines are detected in the optical, they are subject to strong HFS, affecting resulting abundances by up to 0.6~dex if HFS is not properly taken into account \citep{GBS6}. 
Sodium is derived with high precision from optical spectra, but has more problems in the IR. There are three \nai\ lines that are considered largely clean in the optical. Although they are subject to HFS, it seems that the effect is small, since the LLD and MMD for the abundances are small \citep{2014A&A...570A.122S}. In the IR, the two available absorption features are in most cases too weak or blended. Non-LTE effects for \nai\ can be very strong \citep{2011A&A...528A.103L}. Finally, V and Co are elements which are common but more uncertain than the other elements, being subject to HFS. Although V and Co lines are very numerous in the optical, they can be quite weak. They become scarce and too weak in the IR or for metal-poor stars. V and Co abundances are normally derived from blue lines for metal-poor stars \citep{2016ApJ...817...53S}. \\

The fact that our periodic table in Fig.~\ref{fig:periodic_table_surveys} has most cells unmarked means that, for most of the elements, it is either very challenging or impossible to detect and model lines for typical spectra of FGK-type stars in order to obtain accurate abundances. From Fig.~\ref{fig:periodic_table_surveys}, the elements that are potentially detected in less than two surveys correspond to Li, P, S, K, Sc, Zn, Ge, Rb, Sr, Y, Zr, Nb, Mo, Ru, Ba, La, Ce, Pr, Nd, Sm, Eu, Gd, Dy, and Yb. Most of the abundances provided by surveys for these elements are so uncertain that they are not released publicly. Further discussion on these elements can be found in the supplementary text (Sect.~\ref{sup2}). 

\begin{textbox}[b]
\section{Prospects for measuring abundances of new elements in surveys:}
The most challenging elements are those for which the lines are scarce, too weak, and blended. A number of such challenging lines belong to heavy neutron-capture elements, which are part of families of other nucleosynthesis channels than the ones typically measured from survey spectra. They might provide more dimensions in chemical space, and the information might be available for future machine-learning approaches. We must provide a good training set, including the corresponding abundances, using spectra of high resolution, high S/N, and of very extended wavelength range. This is key to maximise the number of industrial abundances that will be extracted from future surveys. 
\end{textbox}

%% file: sect_stellarPops.tex
\cite{2002ARA&A..40..487F} 
discussed the power of using abundances of FGK-type stars in order to find the building blocks of the Galaxy. This is possible if (1) Stars are born in groups which share the same chemical composition; (2) The evolution of this chemical composition is the product of a unique combination of star formation and nucleosynthesis, which depends on where and when the stars were born; and (3) FGK-type stars retain the information of their chemical make-up in their atmospheres. 

In general, there is evidence that the principles mentioned above are correct. 
But when we become ambitious and want to recover the history of each star in the Galaxy, the principles discussed above are challenged by secondary effects in the general picture.  In this section we  discuss a selection of topics where active research is on-going, thanks to the progress in deriving precise abundances for large samples of stars.

\subsection{Nucleosynthesis channels and chemical dimensions}

\cite{2002ARA&A..40..487F} 
indicate that in order to reconstruct the expected $10^8$ star-formation sites in the disk, a chemical ($C$) space of at least ten abundance ratios reflecting different nucleosynthesis channels, at a precision better than 0.05~dex, would be needed. 
It is good news that many current large catalogues contain abundances of more than ten elements (see Figure~\ref{fig:periodic_table_surveys}), but some are correlated due to their similar production mechanisms \citep[see discussion in, e.g.,][]{2016MNRAS.457.3934L}. 
\cite{1957RvMP...29..547B} suggested different chemical families according to their nucleosynthesis paths. 
A detailed description of the nucleosynthesis channels from supernovae can be found in \cite{2013ARA&A..51..457N}. 
In \cite{2014PASA...31...30K}, 
a similar review regarding nucleosynthesis from AGB stars is given. Briefly, elements can be divided into five major families: $\alpha$-capture, iron-peak, odd-Z, light, and neutron-capture. Each of these families contain elements or isotopes which might be produced by different channels (hence environments and timescales!), and so abundance ratios of elements within a family increase the dimensions in $C$ space, serving as diagnostics to study chemical evolution. 

In \cite{2012MNRAS.421.1231T}, 
this dimensionality is studied with a principle component analysis, combining [X/Fe] ratios and using different catalogues from the literature. They found six major components formed by combinations of elements that are correlated with nucleosynthesis channels. Depending on the catalogue (e.g., the amount and precision of abundances) and its overall metallicity distribution, the components are formed by different combinations of abundance ratios. The neutron-capture family was shown to have a large impact on the number of dimensions. Later, \cite{2015ApJ...807..104T} 
continued the discussion, showing that such $C$ spaces would still allow one to find a large number of prominent groups of stars in the Galaxy ($10^{3-4}$). They noted that the number and the size of the groups ($C$ cells) depends more on the uncertainties of the abundances measured than on the number of elements or number of stars that a survey might have. It remains to be seen if considering abundance ratios between or within families might increase the number of dimensions of the $C$ space for chemical tagging with current spectroscopic surveys and abundances uncertainties. For example, it has been shown that populations separate in [Mn/Mg] \citep{2015MNRAS.453..758H},  
[Co/Cr] and [Ca/Mg] \citep{1995AJ....109.2757M}, 
and in [Ba/Eu] and [Ba/$\alpha$] \citep{2009ARA&A..47..371T}. 

\subsection{From chemical tagging to Galactic phylogenetics}

Using the chemical elements to identify the groups of stars that have common origins forms the basis of chemical tagging. As discussed above, this can work if the principle that every group formed at a given place and time in the Universe has a unique chemical pattern. \cite{2002ARA&A..40..487F} 
postulated that this chemical pattern can be attributed to the stellar DNA, such that chemical tagging could allow for temporal sequencing of stars, similar to building a family tree through DNA sequencing. This is only possible because the chemical patterns evolve with time, and not in a random way. In fact, there is a chemodynamical model based on physical principles describing how stellar generations become more metal-rich with time \citep{2006ApJ...653.1145K}. 
The fact that there is a physical process behind the change of the chemical pattern of stars implies that chemical elements carry evolutionary information from one generation of stars to the next. This principle of ancestry forms the basis of phylogenetic studies. In ``Galactic phylogenetics'' \citep{2017MNRAS.467.1140J}, the only useful traits are the chemical elements, because no other trait (kinematics, ages, stellar parameters) carries information that is passed from one generation to the next. Using these traits to construct phylogenetic trees can provide a powerful way to constrain the chemical evolution model underneath, in the same way as many other applications in evolutionary studies.

\subsection{New challenges for chemical evolution with high precision abundances}

As discussed above, improving precision allows one to detect more $C$ cells. With strictly differential techniques, where abundances are derived with respect to reference stars of the same spectral type, precisions of 0.01~dex have been achieved \citep{nissen-gustafsson}. With the help of such precise abundances, it has been shown  that clusters might have chemical inhomogeneities above that level, challenging the principle (1) mentioned at the start of this section. For exampel, \cite{2016MNRAS.457.3934L} 
analysed the Hyades cluster, finding that stars of the cluster can have an abundance dispersion of the order of 0.02 to 0.03~dex. 

For typical uncertainties at a more industrial level, cluster stars have a dispersion in abundances that is of the order of the measurement errors \citep{2016ApJ...817...49B}. 
It is important to quantify this dispersion, as it is key for prospects of chemical tagging. \cite{2016ApJ...833..262H} 
were able to recover stars from open clusters from precise abundances from APOGEE using K-means. However, \cite{2018ApJ...853..198N} 
found a small fraction of field stars that have abundances that are indistinguishable from cluster stars within the uncertainties, yet have different birth origin. Stars of different origins and the same chemical abundances \citep[called doppelg{\"a}ngers by][]{2018ApJ...853..198N} 
should not exist for chemical tagging to work. Accurate ages, kinematics \citep[although see discussions in][]{2014MNRAS.438.2753M}, 
and the inclusion of more dimensions in $C$ space with, e.g., neutron-capture elements, are important for understanding the nature of these stars. 

Atomic diffusion is another challenge. It is well known that heavy elements sink towards the center of stars due to gravitational settling. \cite{2017ApJ...840...99D} 
studied this effect in the context of chemical tagging, showing that abundances can significantly decrease over the lifetime of a star. For the effects of diffusion to be minimised, stars of the same evolutionary stage, and [X/Fe], instead of [X/H] abundance ratios, should be used. It is, however, still plausible that stars of the same evolutionary phase but different masses (and hence ages) will present small differences in [X/Fe] ratios detectable at the 0.01~dex level of precision, which could explain the problem of the doppelg{\"a}ngers, for example. To truly quantify these differences, better theoretical treatment of atomic diffusion is needed, in particular the combined effect with radiative levitation.

High-precision spectroscopic studies have shown that stellar abundances might encode signatures of planet formation. The encoding typically appears as a trend of \xfe\ as a function of condensation temperature ($T_c$). \cite{2009ApJ...704L..66M} 
determined precise abundances of eleven solar twins, and found that the solar refractory elements were more deficient than the volatile elements when compared to other stars. Recently,  \cite{2018arXiv180202576B} 
presented a comprehensive discussion of the chemical homogeneity of sun-like stars considering 79 solar twins with 30 measured elements, finding that the Sun has indeed an unusual slope of $T_c$ vs \xfe. Whether this implies an unusual formation scenario for the planets of the Solar System is still debated. In any case, it is certainly important to keep in mind that there is a possibility that stars which have formed from the same molecular cloud might present different trends of abundances with $T_c$. The binary 16~Cyg AB analysed by \cite{2014ApJ...790L..25T}  
is an example, although the binary $\alpha$~Cen AB analysed by \cite{2018A&A...615A.172M} is a contradictory example. 

\subsection{Masses and ages from stellar abundances}
Recent active discussions in the literature show that masses and ages can be determined from stellar spectra. Such discussions were initiated by \cite{2015MNRAS.453.1855M}, 
who showed that the C/N of APOGEE DR12 red giants revealed that the thin disk and the thick disk had different formation histories because of their different distributions in stellar masses. They explain that C/N relates to mass because of a very fundamental principle in stellar evolution: as red giants experience the dredge-up, synthesised material from the CNO cycle at their cores is brought outwards, which results in an enhancement of the nitrogen surface abundance at the expense of carbon. The amount of mixing depends on the depth of the dredge up, which depends on the mass of the star. 

\cite{2015MNRAS.453.1855M} warn that translating C/N into mass is complicated by the uncertainties in stellar evolutionary models, especially the initial metallicity and C+N abundances, and the poorly understood effects of mixing-length theory, as well as the role of opacities due to $\alpha$-element enhancement and extra mixing in evolved giants. \cite{2015A&A...583A..87S}, \cite{2017MNRAS.464.3021M}, and \cite{2017A&A...601A..27L} 
investigated these complications with dedicated studies of stellar evolution theory and C/N abundances. The complications have, however, not impeded \cite{2016MNRAS.456.3655M} 
or \cite{2018arXiv180409596D} 
from building empirical relations for masses, ages, and C/N thanks to the relative large spectro-seismic datasets for which ages and masses can be derived from asteroseismology.
Such relations have helped to create maps of ages and masses of the Galaxy with data-driven methods \citep[e.g.,][]{2016ApJ...823..114N, 2017ApJ...841...40H, 2018arXiv180409596D},  
with APOGEE or LAMOST data. This is another example of the importance of having seismic data for Galactic studies.

%% file: supplementary_material.tex
\section{SUPPLEMENTAL TEXT: METHODS TO DETERMINE STELLAR PARAMETERS} \label{Supl1}

\subsection{Effective temperature}

\paragraph*{Infrared Flux Method} (IRFM, also called photometric temperatures): If stars were free of atmospheres, they would emit light as a perfect blackbody. Their atmosphere induces some extra absorption, which for FGK-type stars is relatively small at infrared wavelengths \citep{1977MNRAS.180..177B}. This means that, from their photometry at different rather red bands, the near-blackbody model (i.e., the characteristic temperature) which best reproduces the data can be constrained. Based on accurate \teff\ values derived in this way for a calibration sample
it is possible to construct relations of photometric colours with temperature, provided the spectral class and a general estimate of the metallicity are known.  
Widely used relations are those of
\cite{2010A&A...512A..54C}, 
\cite{2005ApJ...626..465R},  
\cite{2009A&A...497..497G}, 
and \cite{1996A&A...313..873A}. 
Since good photometry exists for the large majority of stars with spectra, it is possible to obtain photometric \teff\ values almost for free.
The disadvantage is that the relations depend on extinction. Flux calibration plays another fundamental role  \citep{2010A&A...512A..54C}. Other problems might arise for stars with anomalous abundances (e.g., carbon-enhanced stars) for which the infrared continuous flux might deviate significantly from the simple dependence on \teff\ that lies at the core of the IRFM. It is difficult to identify these stars from photometry alone, which is why it is recommended to validate the photometric  \teff\ values with other methods. Errors in temperature should ideally be derived as the variance of values obtained from relations using different colours, authors, and extinction laws. Typically this variance is at least 100~K. 
When few spectral lines are available (e.g, metal-poor stars, short wavelength coverage, low S/N, or low-resolution spectra) photometric temperatures seem to be the preferred solution. These temperatures are either directly adopted for the programme stars \citep{2012ApJ...753...64I, 2017A&A...604A.129M} 
or used to calibrate the spectroscopic ones \citep[for the APOGEE and RAVE pipelines, respectively, see Sect.~\ref{sect:surveys}]{2016AJ....151..144G, 2017AJ....153...75K}. 

 \paragraph*{Excitation balance:} \fei\ lines are very numerous in spectra of FGK-type stars, and their strength depends on \teff. The \teff\ is chosen such that there is no dependence of the \fei\ abundances obtained on their excitation potentials. This technique is based on the Boltzmann equation, which relates the line strength to the temperature of the layer in the atmosphere where the absorption is produced and the excitation state of the atom. Other atoms which present many lines, like Ti, Ca, or Si, can also be used to determine \teff\ if an insufficient number of Fe lines are available, for example in metal-poor stars. When EWs are measured, the \teff\ is found by removing the slope of the derived abundances as a function of excitation potential. In the synthesis approach, the \teff\ is usually determined by finding the model that provides a good fit to all lines simultaneously.
In either case, excitation-balance temperatures are among the most popular methods due to the large amount of neutral lines found in high-resolution spectra of cool stars, enabling a high internal precision. 
 Typical precisions are better than 50~K. However,  many \fei\ lines suffer from non-LTE effects, so a \teff\ derived with this method has an associated error which is normally neglected. \cite{2014A&A...562A..71B} 
studied solar-type stars, determining \teff\ from \fei\ lines differentially with respect to the Sun. They found that the results changed on average by $\sim$30~K when non-LTE abundance corrections 
were taken into account. 

  \paragraph*{Balmer line profile fitting:} The profiles of \hi\ lines in normalised FGK-type spectra are essentially only affected by \teff. Given the strength of Balmer lines in such spectra, this method has been used extensively over the past 50 years. Accurate \teff\ estimates with Balmer lines are, however, challenging to obtain, since they suffer from modelling and observational uncertainties. The fact that \hi\ lines are so strong means they are formed in deep layers, and so convection becomes important in their modelling. This implies an urgent need to use 3D models for obtaining more reliable profiles. Furthermore, the continuum can only be determined with confidence for high-dispersion observations with  large spectral coverage, which does not occur for most  cross-dispersed echelle spectra. Thanks to the advances in spectral modelling over the last few years, the situation is drastically improving.    \cite{2018AA...615A.139A} 
provide an excellent introduction to the method, where they investigate with 3D-non-LTE  models the temperature determinations of a few \gbs s (Sect.~\ref{sect:gbs}). They improve the accuracy of temperatures by typically 50~K, although for some stars and lines the results can change by up to 200~K. 

 \paragraph*{Interferometry:} Following the Stefan-Boltzmann relation, the bolometric flux, the angular diameter, and the distance of a star can be used to determine its \teff\ from fundamental principles \citep[e.g.,][]{GBS1}. 
These observables can be obtained for the nearest stars, which have large angular sizes on the sky, and for which an angular diameter can be measured with interferometry. Temperatures obtained in this way are perhaps the most accurate ones, with typical uncertainties of about 50~K \citep{2012ApJ...746..101B, 2018MNRAS.475L..81K}. 
However, determining angular diameters is a very difficult task, especially for dwarf stars \citep{2018MNRAS.477.4403W}, 
hence interferometric temperatures are possible for  only a handful of stars. In addition to the observational challenges, the determination of angular diameters require a prescription for the limb-darkening, so angular diameters are not entirely free of modelling assumptions \citep[e.g.,][]{2012A&A...545A..17C, 2018MNRAS.475L..81K}. 

\subsection{Surface gravity}

 \paragraph*{Parallax} (also called trigonometric \logg):
Using the Stefan-Boltzmann relation, the surface gravity can be expressed as a function of mass, \teff, and absolute bolometric magnitude of the star with respect to the Sun \citep[see, e.g., Eq.~1 of ][]{2014AJ....147..136R}. 
\teff\ values are determined by another method (see above), and bolometric magnitudes can be estimated if the distances are known (hence the name of the method). The determination of the mass is more complicated, as it requires the use of evolutionary tracks. Section 4.1 of \cite{GBS1} presents a detailed explanation of this process, and how this propagates into the uncertainties of \logg.  Bolometric magnitudes also become uncertain when the parallax is not accurate, which has  so far been the case for the majority of the stars with spectra.  Because of this, the method has been applied for only a small sample of nearby stary.  For accurate Hipparcos parallaxes, typical internal uncertainties for \logg\  are below 0.1~dex \citep{2004A&A...420..183A}. 
Thanks to the millions of new accurate parallaxes from Gaia, trigonometric \logg\ values will become popular in the next few years. Uncertainties in masses might still challenge the accuracy of the \logg\ values derived in this way.  Fortunately, active research in asteroseismology is rapidly improving stellar evolutionary models, which is leading to better estimates of stellar masses \citep{2017AN....338..644M}. 

 \paragraph*{Ionisation balance:} 
The strength of ionised Fe lines is sensitive to the pressure in the atmosphere. It is thus possible to constrain \logg\ by requiring that the abundance derived from \feii\ lines agrees with that from \fei\ lines. If not enough \feii\ lines are available in the spectrum, other elements can be used, since the principle is the same. This method is widely used, and has typical uncertainties of the order of 0.1~dex. 
The problem with this method is that, in some cases, very few ionised lines are available. \cite{2013AJ....145...13A} 
discuss the lack of \feii\ lines in  metal-poor turn-off  stars, which forced these authors to set the gravity to a fixed value of \logg$= 4.0$, with conservative uncertainties of 0.5 dex. 
In addition, 
aiming for ionisation balance might mean adopting \logg\ values that are not physical if overionisation due to non-LTE is strong. 

 \paragraph*{Wings of strong lines:} The absorption in the wings of strong lines such as the \mgi~b or the \caii\ triplet around $\lambda$5184\AA\ and $\lambda$8541\AA, respectively, is also sensitive to the pressure in the atmosphere. Since these are metallic lines, in order to determine \logg, the metallicity should be known independently. Note that these lines correspond to the absorption of $\alpha$-elements, hence one needs to define \afe\ in the model as well (see Sect.~\ref{sect:otherparams}). The advantage of the method of fitting synthetic to observed strong-line profiles is that these features are strong in all FGK-type stars, making it possible to determine \logg\ from spectra with lower resolution. 
The sensitivity  of the wings is, however, not sufficient to detect variations of 0.1~dex or less in \logg\ if the S/N is low, which is why this parameter cannot be constrained to within 0.2~dex or below for some stars.

 \paragraph*{Asteroseismology:} 
A relatively new way to determine surface gravities is using asteroseismology. As discussed in the review of \cite{2013ARA&A..51..353C}, 
frequency modes of solar-like oscillations can be directly related to \logg\ and \teff. These scaling relations allow one to obtain very accurate and precise \logg\ values for a given \teff. Uncertainties can be as small as 0.01~dex. Frequency modes can be determined from data obtained by space missions such as {\it Kepler} \citep{2010Sci...327..977B}, K2 \citep{2014PASP..126..398H} and {\it CoRoT} \citep{2008Sci...322..558M}.  As shown by the APOKASC campaign  \citep{2014ApJS..215...19P},  
seismic detections are helping to calibrate not only the parameters in spectroscopic surveys, but also stellar models for Milky Way science in general \citep{2016A&A...594A..43H, 2017A&A...600A..66V, 2017ApJS..233...23S}. 
However, it is still difficult to obtain good asteroseismic data for metal-poor stars, which are usually faint. Surface gravities for such stars will still be problematic to constrain in the next decades, but some efforts are on-going \citep{2018arXiv180808569V}. 
\cite{2017AN....338..644M} 
present a recent discussion of how asteroseismology is contributing to the field, in particular, in the determination of ages and improvement of spectroscopic parameters.

\subsection{Metallicity}

\paragraph*{Iron lines:}
The determination of \feh\ from a cross-dispersed high-resolution echelle spectrum normally will be done by measuring the strength of iron lines. The reason for using iron for an overall metallicity estimate is that the number of \fei\ and \feii\ lines in the spectra of FGK-type stars is by far the largest among all elements. Hence, the precision with which Fe abundances can be measured is the highest. Often iron abundances from neutral and ionised lines do not agree (e.g., ionisation imbalance), which may be attributed to uncertainties in 1D-LTE modelling \citep[see discussions in, e.g.,][]{2004A&A...420..183A, 2012MNRAS.427...27B, 2017ApJ...847..142E}. 
Since, in general, there are more \fei\ than \feii\ lines available, [\fei/H] is more precise than [\feii/H]. 
However, since for many stars \fei\ cannot be accurately modelled in LTE, and low-excitation lines are subject to 3D effects, even though the \feii\ results are less precise, they might be more accurate. If ionisation balance can not be restored, a decision needs to be made on how the \fei\ and \feii\ results are combined into a final metallicity value. For example, in \cite{GBS3} the non-LTE corrected \fei\ value was used to represent metallicity, adding as an uncertainty the difference with \feii\ abundances.  
As stellar abundances in general, metallicities are affected by line selection, atomic data, normalisation, the method employed to analyse the lines (synthesis or EW), and the temperature and surface gravity scale.  This is reflected by a median uncertainty of 0.06~dex when considering the different \feh\ values reported in the literature for common stars in PASTEL \citep{2016A&A...591A.118S}.

\paragraph*{Global spectral fitting of data to a grid of synthetic spectra:}
This is the only way to estimate metallicity when few or no iron lines are detectable, which is the case for lower resolution spectra.  This way to determine metallicities can also be applied to high-resolution spectra with extended wavelength coverage if  a representative grid of synthetic spectra is provided. 
The metallicity from APOGEE spectra, for example, is determined from several molecular features, in addition to iron and $\alpha$-element lines \citep{2016AJ....151..144G}. 
Another example is the AMBRE project \citep{2012A&A...544A.126D},  
in which parameters of the HARPS, FEROS and UVES spectra in the ESO archive are determined \citep[][and references therein]{2016A&A...591A..81W}. 
Metallicities in these cases are normally reported as \mh. 
To quantify the uncertainties of metallicities obtained with these methods, we use comparisons of \mh\ between  stars in common for  APOGEE and LAMOST by \cite{2018arXiv180707625A}, 
who report a scatter of 0.13~dex in this parameter. 

\section{SUPPLEMENTAL TEXT: DISCUSSION ON COMPLICATED ELEMENTS}\label{sup2} 

We base our discussion on the information reported by \citet[the GES line list]{heiter-linelist}, \citet[Hypatia]{2014AJ....148...54H}, and \citet[GES-UVES]{2014A&A...570A.122S} for the situation in the optical, and on \cite{2018arXiv180709784J} and \cite{2016A&A...594A..43H} for the situation in the IR.  

In Hypatia, Sc, Zn, Y, and Ba have abundances reported for more than half of the stars (see Fig.~\ref{4.hypatia}). They are popular because they present good clean lines, although very few, and only in the optical. We note, however, that Sc, Zn, and Y have a large MMD in GES. It remains to be seen whether LAMOST-like spectra would present signatures of such elements, provided that data-driven methods were trained on optical spectra. Li, S, Sr, Zr, La, Ce, Nd, Sm, and Eu are reported in Hypatia for less than half of the sample but more than 1000 stars. As in the previous case, these elements present few clean lines only, except for S, Ce, and Nd. S presents several clean features in the IR, making it a precise element in APOGEE, but in the optical this element is more uncertain, with one clean line with good laboratory data only. We include Ce and Nd in this discussion despite being marked by more than two surveys in Fig.~\ref{fig:periodic_table_surveys}. This is because new identifications of \ceii\ and \ndii\ lines in APOGEE spectra have been reported by \cite{2017ApJ...844..145C} and \cite{2016ApJ...833...81H}. In the optical, twelve \ceii\ lines are detected, among which two are free of blends and have good atomic data and can be used for reliable abundance determination. Over fifty optical \ndii\ lines are detected, however, they are subject to hyperfine and isotope splitting, and most of them are very blended.

Elemental abundances that are reported in Hypatia for less than 1000 stars are P, K, Nb, Mo, Ru, Gd, Dy, and Yb. Phosphorus presents three lines in the IR, but they are blended and very weak. The 21 stars with P abundances in Hypatia come from the analysis of IR lines or from UV lines for metal-poor stars. Potassium can be detected in visual spectra, but in a region that is very contaminated with telluric absorption \citep{2017A&A...600A.104M}. 
There are two K lines in the IR that are slightly blended with CN molecules, but otherwise have good strength for abundance determination for large samples of stars. Since Nb, Gd, and Dy are included in the GES line list, we have marked them in Fig.~\ref{fig:periodic_table_surveys}. However, the nine weak \nbi\ lines are heavily blended. Gd and Dy have one identified line each in GES spectra, but both are very weak and blended in FGK-type stars. The two suitable optical Mo lines are too weak in dwarfs to be useful and Ru has so far only one good line which is partially blended. Although Ru lines are subject to HFS and isotope splitting, because the line is weak these effects are expected to be negligible. While not included in GALAH DR2, these Mo and Ru lines fall within its wavelength coverage and may be included in future data releases \citep{2018MNRAS.478.4513B}. 

Other elements which can potentially be analysed in optical and IR spectra are rubidium and ytterbium. The \rbi\ line at $\lambda$7800\AA\ has been analysed by \cite{2014AJ....147..136R} 
to provide abundances for metal-poor stars and is included in the GALAH spectra. The IR \rbi\ $\lambda$15290\AA\ line is very weak and heavily blended, presenting a challenge for most of the stars observed by APOGEE. In the optical there is an \ybii\ line at $\lambda$3694\AA\ which can be used for analysing metal-poor stars \citep{2014AJ....147..136R}, 
but it becomes too blended for more metal-rich stars. In the IR, a weak and blended \ybii\ line at $\lambda$16498\AA\ has been identified and could be included in future data releases of APOGEE, provided the syntheses around the blends are accurate. Finally, Ge has the potential to be detected in IR spectra as well. The \gei\ line at $\lambda$16760\AA\ is heavily blended with an \fei\ line at the resolution of APOGEE, but could be resolved at higher resolution.

We finish this section discussing the elements for which abundances are included in Hypatia, but which have no record of detection in spectroscopic surveys. These elements correspond to Be, Pd, Ag, Tb, Er, Tm, Hf, Pb, and Th, and are enclosed with a magenta square in Fig.~\ref{fig:periodic_table_surveys}. Their abundances have been reported in the literature from spectral analyses of FGK-type stars, but mainly from a few lines detected towards the blue end of the visual spectrum \citep{2014AJ....147..136R}. Thus, if nucleosynthesis channels like the rapid neutron-capture process are to be explored in future spectroscopic surveys, it is important to extend the wavelength coverage towards the blue side of the spectrum (see \citealt{2015AN....336..665H} for a discussion for 4MOST). 